\documentclass[aps,pre,notitlepage,unsortedaddress]{revtex4-2}
\usepackage{amssymb,amsmath,bm}
\usepackage{graphicx}
\usepackage{enumerate}
\usepackage{color}
\usepackage{hyperref}

\begin{document}

\title{Metropolis Monte Carlo sampling: convergence, localization transition and optimality}

\author{Alexei D. Chepelianskii}
\affiliation{Universit\'e Paris-Saclay, CNRS, Laboratoire de Physique des Solides, 91405 Orsay, France.}
\author{Satya N. Majumdar}
\affiliation{Universit\'e Paris-Saclay, CNRS, LPTMS, 91405 Orsay, France.}
\author{Hendrik Schawe}
\affiliation{LPTM, UMR 8089, CY Cergy Paris Universit\'e, CNRS, 95000 Cergy, France.}
\author{Emmanuel Trizac}
\affiliation{Universit\'e Paris-Saclay, CNRS, LPTMS, 91405 Orsay, France.}

\begin{abstract}
Among random sampling methods, Markov Chain Monte Carlo algorithms are foremost.
Using a combination of analytical and numerical approaches, we study their convergence properties towards the steady state,
within a random walk Metropolis scheme.
Analysing the relaxation properties of some model algorithms sufficiently simple to enable analytic progress, we show that the deviations from the target steady-state distribution can feature a localization transition as a function of the characteristic length of the attempted jumps defining the random walk.
While the iteration of the Monte Carlo algorithm converges to equilibrium for all choices of jump parameters, 
the localization transition changes drastically the asymptotic shape of the difference between the probability distribution reached after a finite number of steps of the algorithm and the target equilibrium distribution.
We argue that the relaxation before and after the localisation transition is respectively limited by diffusion and rejection rates. 
\end{abstract}

\maketitle

\section{Introduction} 

Although Buffon's needle problem \cite{Buffon} 
may be considered as the earliest documented use of Monte Carlo sampling (18th century),
the method was developed at the end of the second world war and dates from the early days of computer use \cite{Metropolis_Ulam,eckhardt1987stan}. 
With the increase in computational power, it 
has become a pervasive and versatile technique in basic sciences and 
engineering. It uses random 
sampling for solving both deterministic and stochastic problems,
as found in physics, biology, chemistry, or artificial intelligence 
\cite{mcbook,Binder_book,mode2011applications,bolhuis2002transition,becca2017quantum,bishop2006pattern,Shaebani2020}.
Monte Carlo techniques also allow to assess risk in quantitative analysis and decision making \cite{Rubinstein,Glasserman2003},
and their methodological developments provide tools for economy, epidemiology
or archaeology \cite{Gilks}.
It is then
crucial to understand the type of errors which can be introduced as a consequence of the incomplete convergence of such algorithms.

Our interest goes to Markov Chain Monte Carlo techniques \cite{Roberts,NewmanBarkema},
that create
correlated random samples from a target distribution;
a special emphasis is put on the relaxation rate of these methods.
From the target probability distribution, 
a sequence of samples is obtained
by a random walk, with appropriate transition probabilities. 
The walker's density evolves at long times towards the target distribution,
and quantities of interest follow from the law of large numbers
and other methods of statistical inference \cite{Rubinstein,Glasserman2003,Gilks,Roberts,FrenkelSmith,Krauth,NewmanBarkema,Wasserman,Bedard}.
The Monte-Carlo method and its modern developments \cite{Duane1987,Creutz1989,wolff,Wong2000,Roberts2001,Bernard2009event,Hsu2011,Michel2014generalized,Krauth2017,Volker2019,QCD2020} have now been adopted in many topics in and outside physics. %, yet no exact results on the relaxation rate  
%Rejection based approaches on the other hand appear more flexible and arguably simpler,
%which explains why they remain so broadly used.}
A key issue deals with the speed of convergence of the algorithm: the larger
the convergence time, the larger the error bars for the computed quantities. 
For most practical applications, % instead of \cite{rque1} 
if the amplitude $a$ of the random jumps is small, phase space 
is not sufficiently explored, even though most attempted jumps are accepted.
%sampling is inefficient although most attempted jumps are accepted. 
Conversely, large jumps will lead to a
large rejection probability, and to an equally ineffective method, see Appendix \ref{sec:optimal}. 
In between, one expects an optimal jump size $a_{\text{opt}}$ at which the convergence rate is maximal. There have been theoretical attempts in deriving
$a_{\text{opt}}$ for specific models \cite{Gelman}. In practice, without a precise knowledge of $a_{\text{opt}}$, a widely accepted rule of thumb is to choose $a$ such that 
the acceptance probability is close to 50\% for the attempted moves
\cite{Krauth,Allen,FrenkelSmith,Talbot}. 
%Beyond its optimality, the physical significance of $a_{\text{opt}}$ has not been explored.
In this paper, we show the existence of a new critical value $a^*$,
where an unexpected localization transition occurs such that 
the relaxation mechanism is drastically different 
for $a<a^*$ and $a>a^*$.
%We show that as $a$ is varied, an unexpected {\em localization} transition, where
%the nature of the error changes strikingly, 
%takes place generically at a critical value 
%at $a=a^*$. 
%Our results demonstrate that %near $a^*$, 
%In coincides, rather remarkably, precisely with  
%when it is present, $a_{\text{opt}}$.
%Thus, near the optimal convergence rate, 
%a small change in algorithmic parameters
This
deeply modifies the nature and amplitude of the error. 
The existence of the critical value $a^*$ is our main finding,
and the novelty of this work. Besides,
although there is no reason to expect any relation between 
$a_{\text{opt}}$ and $a^*$, we report, rather interestingly, 
a number of examples, where they coincide precisely.
%or are very close to each other.
We emphasize that while a number of results have been proven for relaxation rates, those mainly hold close to the diffusive limit \cite{Dey2019,Jourdain2015,Peskun1973}; there, powerful mathematical techniques based on micro-local analysis have been developed \cite{Diaconis2009,Diaconis2011,Diaconis2012}, leading to accurate results on the relaxation rates. In the present work, we explore a regime well beyond the diffusive limit, which has so far been investigated only through numerical simulations \cite{Diaconis}. As an alternative approach, we focus on the study of the relaxation eigenmodes, which allows us to obtain accurate analytic results for relaxation rates, valid all the way up to the localization transition, thus far from the small jump diffusive region. We also obtain a scaling function description of the relaxation in the localizing phase.

The paper is organized as follows. In section \ref{sec:2_master}, the formalism is laid, 
with the Master equation approach. Section \ref{sec:relax} contains our main findings,
with emphasis on relaxation to equilibrium, and the localization transition for the leading relaxation eigenvectors of this Master equation. In section \ref{sec:4_analyt},  we show, how to construct accurate analytic approximations to the relaxation rate before the localization onset, using the Fokker-Planck limit eigenvectors of the Master equation. 
Our conclusion is presented in section \ref{sec:5_concl}.
For the ease of reading,
more technical developments are relegated to six appendices. We provide a derivation of the Master equation and introduce the analytical tools that are used for its investigation (see Appendices \ref{sec:optimal}, \ref{sec:master} and \ref{sec:overview}).  In Appendix \ref{sec:box}, we show analytically that in the localized phase, relaxation eigenvectors are replaced by a self-similar relaxation ansatz. In Appendix \ref{sec:analytical}, we show how the relaxation rate can be computed (semi)analytically using the Fokker-Planck eigenvector basis.
In this paper, we have chosen to focus mainly on specific 1D cases for which exact (or highly accurate approximate) analytical treatement was possible. In Appendix \ref{sec:generalizations}, we present additional numerical evidence for a localization transition with more general 1D and higher dimensional examples leaving the analytic treatment of higher dimensional cases for future works. 

\section{Master equation for Metropolis Monte-Carlo sampling and relaxation to equilibrium} 
\label{sec:2_master}

We start with by reminding general results on the Master equation describing relaxation of the Metropolis Monte-Carlo algorithm. The spectral properties of Markov-Chains have been extensively studied in the Mathematical literature \cite{Levin,Douc}. Here we instead focus on the nature of the eigenvectors which have received much less attention.
This introduction will allow us to fix notations and to contrast the relaxation of a discrete Markov-chain with the relaxation properties that we find for the continuous case.

The Markov Chain Monte Carlo method amounts to considering
a random walker with position $x$ (here on the line), in the presence of an external confining potential $U(x)$.
We adopt the framework of the Metropolis algorithm \cite{Metropolis,Hastings,FrenkelSmith,NewmanBarkema,Krauth}.
The position of the particle evolves in discrete time steps $n$ following the 
rule
\begin{eqnarray}
x_n= \begin{cases}
x_{n-1} +\eta_n & {\rm with}\,\, {\rm prob.}\,\,\, p= {\rm min}\left(1, e^{-\beta\, \Delta U}\right)\quad \\
\\
x_{n-1} & {\rm with}\,\, {\rm prob.}\,\,\, 1-p\, ,
\end{cases}
\label{eq:Metropolis}
\end{eqnarray}
where $\Delta U=U(x_{n-1} +\eta_n)-U(x_{n-1})$ and $\beta=1/(k_BT)$ denotes inverse temperature. The random jumps $\eta_n$ at different times are independent,
drawn from a continuous and symmetric probability distribution $w(\eta)$. 
In other words, the particle attempts at time $n$ a displacement $\eta_n$ from its current location $x_{n-1}$, which is
definitely accepted (with probability $1$) if it leads to an energy decrease,  but is accepted with a lesser probability $e^{-\beta \Delta U}$ if the move leads to an energy increase $\Delta U>0$.
A key quantity in what follows is the amplitude $a$ of the attempted jumps,
that we introduce as the characteristic length associated with $w(\eta)$, taken to obey the 
scaling form 
\begin{equation}
    w(\eta) \,=\, \frac{1}{a} f\left(\eta / a\right) .
    \label{eq:w_f}
\end{equation}
Normalization demands that $\int f=\int w =1$.

The dynamics encoded in Eq.~\eqref{eq:Metropolis} can be written in terms of a Master equation for $P_n(x)$, the probability density of the walker at time $n$,
\begin{equation}
P_n(x)= \int_{-\infty}^{\infty} F_\beta(x,y)\, P_{n-1}(y)\, dy .
\label{eq:master}
\end{equation}
The explicit form of the temperature-dependent 
kernel $F_\beta$ is given below in Eq.~(\ref{eq:Fbeta}). Generically, $P_n(x)$
converges towards the target distribution \cite{Hill,bapat}, given by the (equilibrium) Gibbs-Boltzmann expression $P_{\infty}(x) \propto  \exp(-\beta\, U(x))$, see Appendix \ref{sec:master}. We assume that $U$
is confining enough so that $ \exp(-\beta\, U(x))$ is integrable, and for simplicity that $U(x)=U(-x)$. 
Our main interest is to find how quickly the dynamics converges towards the
target density, and with which error $\delta P_n(x) = P_n(x)-P_\infty(x)$.
The convergence rate can be defined from the large time limit of the deviation from equilibrium of some observable ${\cal O}(x)$:
\begin{equation}
\log\Lambda  \,=\, \sup_{\{{\cal O}(x),P_0(x)\}}  \lim_{n \to \infty} \frac{1}{n} \log \left| \int {\cal O}(x) \delta P_n(x) dx \right| 
\label{eq:Lambda_def}
\end{equation}
where the maximum is taken over all possible smooth and sufficiently localized functions ${\cal O}(x)$ and initial distributions $P_0(x)$ which allow numerical estimation.
If $\Lambda < 1$ (given that $\Lambda\leq 1$) the probability distribution $P_n(x)$ converges exponentially fast to the equilibrium distribution for large $n$, i.e., $|P_n(x)- P_\infty(x)| \propto \Lambda^n \propto e^{-n/\tau}$ where
$\tau = - (\log \Lambda)^{-1} $
denotes the convergence time (in number of Monte-Carlo algorithm steps unit).
The convergence rate $-\log(\Lambda) >0$ is the figure of merit of the algorithm; the smaller the
$\Lambda$, the larger the rate,
the smaller the convergence time and the more efficient the sampling is.

The relaxation properties of discrete Markov-chains are well established mathematically, and we will now discuss the connection between our definition of the convergence rate and the relaxation quantities that are used in the Mathematical literature. Two main quantities are introduced to characterise relaxation in this context \cite{Levin}. The first quantity, the mixing time, describes the number of steps required for the probability distribution $P_n(x)$ to deviate less than $\epsilon>0$ from $P_\infty(x)$ where the total variation distance is used as the distance metric. This quantity explicitly depends on the target precision $\epsilon$ and our formal definition for $\Lambda$ can be viewed as the leading asymptotic behavior of this quantity for $\epsilon \rightarrow 0$.  The second quantity, the relaxation time, is defined from the eigenvalue spectrum of the Master equation. Taking $\lambda_r$ as the eigenvalue with largest modulus $<1$, the relaxation time is then defined as the inverse spectral gap $1/(1 - |\lambda_r|)$. The Levin-Peres-Wilmer theorem, chapter 12 in \cite{Levin}, establishes a connection between the two quantities, the relaxation time providing the leading asymptotic behavior for the mixing time in the $\epsilon \rightarrow 0$ limit. To our knowledge, there is no generalization of such a theorem to infinite Markov-chains. The spectral theorem, implies that $1-\Lambda$ will coincide with the spectral gap of the master equation if both the observables functions ${\cal O}(x)$ and the initial probability distributions $P_0(x)$ are all in $L^2(P_\infty)$. This is not exactly our case as we consider, for example, the case of $\delta$ function like initial distributions $P_0(x)$ localized at a single point and those are not in $L^2(P_\infty)$. Our numerical simulations suggest that the relaxation rates obtained from different numerical methods are all consistent (direct Monte Carlo simulations, Master equation diagonalization or forward iteration of the Master equation) and thus it seems safe to think that $1-\Lambda$ coincides with the spectral gap of the Master equation (although we will see that eigendecomposition will differ from what we know for the Schr\"odinger equation). 
To conclude this discussion on the  definition of the characteristic Markov-chain relaxation times we mention that we found it preferable to work with $\Lambda$ directly instead of the inverse spectral gap, as $\Lambda \rightarrow 1$ in the limit of small jumps sizes in the Metropolis-algorithm, while the inverse spectral gap diverges.

A quantity of central importance in the approach is the rejection probability $R(x)$ (see Appendix \ref{sec:master}), or more precisely the fraction of rejected moves per attempted jump:
% \footnote{Although $R(x)$, strictly speaking, is a probability and lies between 0 and 1, it is referred to as the rejection probability, since it counts the fraction of rejected moves, per attempted jump.}
%\begin{eqnarray}
%    R(x)  = \int_{-\infty}^{\infty} dy\, w(y-x) & \left(1 - e^{-\beta \, \left(U(y)- U(x)\right)} \right) \, \nonumber \\
%    &    \theta\left(U(y)-U(x)\right) , \label{eq:rejection_rate}
%\end{eqnarray}
\begin{equation}
    R(x)  = \int_{-\infty}^{\infty} dy\, w(y-x)  \left(1 - e^{-\beta \, \left(U(y)- U(x)\right)} \right) \,  \theta\left(U(y)-U(x)\right) , \label{eq:rejection_rate}
\end{equation}
 %\begin{align}
 %R(x) = \int_{U(y)>U(x)} dy \; w(y-x) \left(1 - e^{-\beta \, \left(U(y)- U(x)\right)} %\right)
 %   \end{align}
where $\theta(z)$ is the Heaviside function: $\theta(z)=1$ for $z>  0$ and 
$\theta(z)=0$ for $z<0$. 
Thus, the rejection probability $R(x)$ from the current location $x$ is zero  if the new position $y$ occurs downhill. 
%In the opposite case of an uphill move, it is $1- \langle e^{-\beta \Delta U}\rangle$ where the average is over all attempted jump lengths.
The integral kernel of the Master equation is then given by (see Appendix \ref{sec:master})
%\begin{equation}
\begin{align}
F_\beta(x,y) &=  \delta(x-y) R(x) + w(x-y) \bigl[ \theta\left(U(y)-U(x)\right) +e^{-\beta\,\left(U(x)- U(y)\right)}\, \theta\left(U(x)-U(y)\right) \bigr]
\label{eq:Fbeta} .
\end{align}
%\end{equation}
Averaging over the position of the particle yields 
the mean rejection probability $R_n= \int R(x)\, P_n(x)\, dx$ which is monitored by default
in all rejection-based algorithms.
This is the quantity that the practitioner aims at keeping close to 50\%,
following a  time honored rule of thumb stating that this provides efficient sampling \cite{Krauth,Allen}.
In the limit $n\to \infty$, this mean rejection probability approaches the stationary value $R_\infty$.
Rigorous studies, in a one-dimensional harmonically confined setting
with a Gaussian jump distribution, have found that the optimal acceptance probability $1-R_\infty$ is close to 44\%, while this quantity may decay when increasing space dimension \cite{Gelman}.
On intuitive grounds, one may expect a relation between $R(x)$ and the 
convergence rate of the algorithm. Indeed, starting from an arbitrary 
point $x_0$ at time $n=0$, the density at time $n$, given $x_0$, can be written
\begin{equation}
    P_n(x|x_0) = R(x_0)^n \delta(x-x_0) + {p}_n(x|x_0)
\end{equation}
where ${p}_n(x|x_0)$ is a smooth function. Thus, an observable $\cal O$
that would only measure the walker's presence in the immediate vicinity of $x_0$, for instance ${\cal O}_n(x_0) = {\rm lim}_{\epsilon \rightarrow 0} \int_{x_0 - \epsilon}^{x_0 + \epsilon} P_n(x) dx$,
% \footnote{For instance, ${\cal O}_n(x_0) = {\rm lim}_{\epsilon \rightarrow 0} \int_{x_0 - \epsilon}^{x_0 + \epsilon} P_n(x) dx$}  
would decay as $R(x_0)^n$. The system as a whole cannot relax faster,
and we obtain from Eq.~\eqref{eq:Lambda_def} a lower bound for the convergence rate,
corresponding to $\Lambda>R(x_0)$, which holds for all choices of $x_0$:
\begin{align}
\Lambda \ge \max_{x_0} R(x_0) .
\label{eq:convergence_rate}
\end{align}
Our objective is to study $\Lambda$ as a function of $a$, for a fixed choice 
of $U(x)$ and $f(z)$. We expect $\Lambda$ to be minimum at a well defined value $a=a_{\text{opt}}$.

More precisely, for a given confining potential $U(x)$ and type of jumps
$f(z)$,
the convergence rate and the resulting error are encoded 
in the spectral properties of the kernel $F_\beta(x,y)$ in Eq.~\eqref{eq:master}. 
We have attacked
this question by four complementary techniques: the derivation of 
exact results,
%(general analytical bounds, together with explicit solutions in specific cases)
numerical diagonalization, numerical iteration of the Master equation, and direct Monte Carlo simulation of the 
random walk dynamics, with proper averaging over multiple realizations
to gather statistics, see Appendix \ref{sec:overview}. We begin with a 
{\em discretized} approximation to the Master equation \eqref{eq:master}, for which Perron-Frobenius theorem shows that the equilibrium state, reached at large $n$ (formally $n\to \infty$), is unique \cite{bapat}: it is given
by  $P_{\infty}(x)$.
At any time, the probability density can furthermore be decomposed 
as 
\begin{equation}
    P_n(x) \, =\, \sum_\lambda {\cal A}_\lambda \,{\cal P}_\lambda(x) \, \lambda^n 
    \label{eq:eigen_discrete}
\end{equation}
where 
the eigenvectors of $F_\beta$ are denoted by ${\cal P}_\lambda(x)$, and
the eigenvalues $\lambda$ can be proven to be real \cite{Levin}, see also Appendix \ref{sec:master}.
Indeed, detailed balance \cite{FrenkelSmith,Krauth,NewmanBarkema}
allows to transform the Master equation into a self-adjoint problem, similarly to the mapping between the Fokker-Planck and Schr\"odinger 
equations \cite{Risken}. 
The precise form of the projection coefficients ${\cal A}_\lambda$ is not essential.
%, but it can be noted that since the dynamics converges to $P_\infty(x)$, 
%${\cal A}_{\lambda_0} =1$. 
%\emma{
Ordering eigenvalues in decreasing order
($\lambda_0 > \lambda_1 \geq \lambda_2\ldots$),
the eigenvalue $\lambda_0=1$ is associated with equilibrium, with eigenvector 
$P_\infty(x)$. For all the cases considered here the modulus of the negative eigenvalues is $<\lambda_1$, thus the 
asymptotic error $\delta P_n$ behaves like ${\cal P}_{\lambda_1}(x)$,
and decays to 0 like $\lambda_1^n$ (also meaning that $\Lambda=\lambda_1$).
Finding the optimal $a$ is a minmax problem,
where one should minimize $\Lambda=\lambda_1$, i.e.\ the maximum eigenvalue,
leaving aside the top (equilibrium) eigenvalue $\lambda_0=1$.

\begin{figure}[tb]
\includegraphics[width=0.5\textwidth]{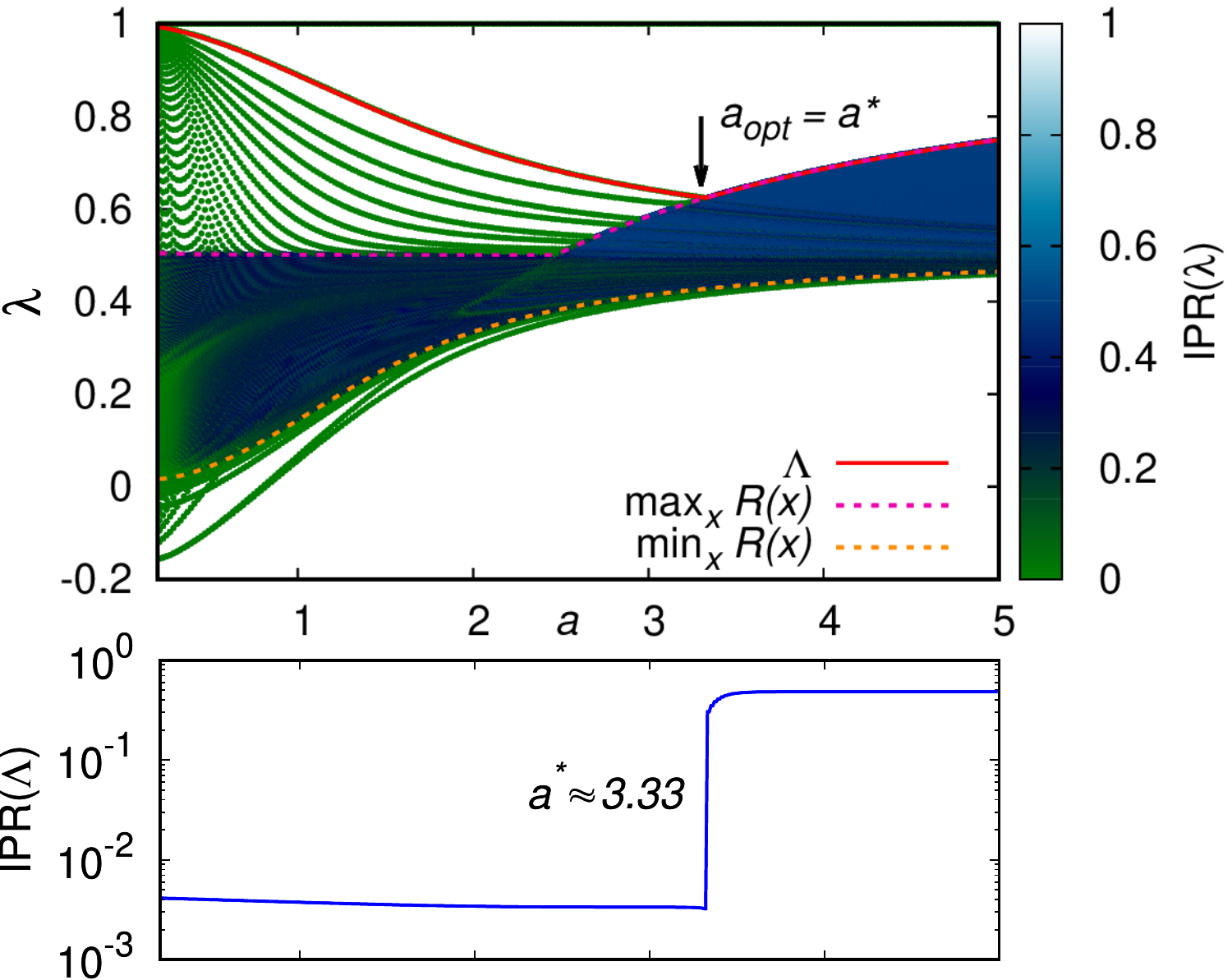}
\caption{The top panel shows the spectrum of $F_\beta$ for harmonic confinement $U(x) = x^2/2$ as a function of jump amplitude $a$,
%Top (gapped situation): Box confinement ($U=0$ in a finite box, while $U$ is infinite outside the box) with exponential jump distribution ($f(x)=e^{-|x|}/2$). The dominant mode, the $\lambda_1$ branch is always above the singular continuum (i.e. ${\cal N} >1$ for all $a$).
%Bottom (gaples situation): 
for a uniform jump distribution of range $(-a,a)$. The color code, provided on the right-hand-side, is for the 
Inverse Participation Ratio of the eigenvector associated to the eigenvalue displayed (see Appendix \ref{sec:overview}).
The upper envelope of the relaxation spectrum defines $\Lambda$, see Eq.~\eqref{eq:Lambda_def}; shown by the red line, it reaches its minimum for $a=a_{\text{opt}} \simeq 3.33$.
This value coincides with the threshold $a^*$ for localization. 
Here, $\cal N$, denoting the number of discrete relaxation modes
(excluding the stationary state), is $0$ for large jumps ($a>a^*$), 
while ${\cal N}$ quickly grows as $a$ diminishes.
Dashed lines show the bounds for the singular continuum,
that appears in dark blue, see Eqs. 
\eqref{eq:rejection_rate} and \eqref{eq:convergence_rate}.
The bottom panel is for the IPR %Inverse Participation Ratio
associated to $\Lambda$, as a function of $a$
(same abscissa as the upper panel).
The localization transition is signaled by the sharp jump %of the IPR 
at $a=a^*$. This threshold does not depend on the number of sites $N_d$, as long as $N_d$ is large enough. Here $N_d=1000$.
The length unit is the thermal length, meaning the standard deviation 
of $P_\infty(x)$.
}
\label{fig:SpectrumBoxExp}
\end{figure}

\section{Relaxation to equilibrium and localization} 
\label{sec:relax}

While the above results hold for the discretized version of
Eq.~\eqref{eq:master}, explicit analytical calculations of the spectrum 
for a number of potentials $U(x)$ reveal that the eigenvector decomposition 
\eqref{eq:eigen_discrete} fails in the continuum limit. In addition 
to the discrete spectrum with well defined eigenfunctions, a 
continuum of eigenvalues appears, with singular localized 
eigenfunctions which in the continuum limit collapse to a point $x_0$ where they take
a finite value. The corresponding eigenvalue is $R(x_0)$. The continuum of these eigenvalues is very different from the continuum spectrum of the Schr\"odinger equation for which eigenfunctions are smooth delocalized functions which extend all the way to infinity with non zero $L^2$ norm. To emphasise the difference with the Schr\"odinger equation continuum, we call this continuum of eigenvalues the singular spectrum. The singular continuum is therefore bounded from below and above by $\min_{x} R(x)$ and $\max_{x} R(x)$. 

Equation \eqref{eq:eigen_discrete} now takes the form
\begin{equation}
   P_n(x) \, =\, \sum_{\lambda \in \{\lambda_{0} \ldots \lambda_{{\cal N}} \}} {\cal A}_\lambda \,{\cal P}_\lambda(x) \, \lambda^n  \,+\, {\cal L}_n(x) ,
    \label{eq:eigen_discrete_continuous}  
\end{equation}
where ${\cal L}_n(x)$ stems from the singular continuum.
Here, the discrete summation runs over a finite (and possibly small) number of $1+\cal N$ terms:
${\cal N}\geq 0$ since the term $\lambda_0=1$ is necessarily present in the
expansion, to ensure the proper steady state. 
The remaining term ${\cal L}_n(x)$ localizes at large times $n\to \infty$ around a finite number of points $x_l$ where the rejection 
rate $R(x)$ in \eqref{eq:rejection_rate} is maximal: $\lim_{n \rightarrow \infty} {\cal L}_n(x)/{\cal L}_n(x_l) = 0$ for any $x \ne x_l$. 
This property of the localizing term ${\cal L}_n$ is valid only for the non-discretized Master equation and is thus most directly established by analytical means. From our analytical computations, two possible scenarios emerge:
(i) ${\cal N} > 0$ for all $a$ and (ii) ${\cal N} = 0$ for $a > a^*$ where $a^*$ gives the position for the localization transition; 
$a^*$ marks the transition from a diffusion governed 
evolution to a phase where relaxation is limited by rejected moves.
In case (i),
the eigenvalue $\lambda_1$ lies above the singular continuum and $\Lambda=\lambda_1$. The error is ruled by a ``regular'' eigenmode akin to what would be found in the discretized approximation. 
In case (ii) on the contrary, 
$\lambda_1$ merges with the singular continuum
at $a=a^*$ and the error is dominated by the localizing term ${\cal L}_n(x)$. Numerical simulations suggest that this localized scenario (ii) is the  generic case,
see also Appendix \ref{sec:box}.
In Fig.~\ref{fig:SpectrumBoxExp}, we illustrate the merging between regular and singular spectrum for the harmonic potential with a flat jump distribution.
To distinguish numerically the regular spectrum as in Eq.~\eqref{eq:eigen_discrete} from the singular one,
we have discretized $F_\beta(x,x')$ into a matrix of size $N_d\times N_d$, and computed the spectrum. 
Two methods have then been employed, both relying on a large $N_d$ analysis.
For the regular part, the spacing between successive eigenvalues stay non-zero as $N_d\to \infty$ while they do vanish in the singular part.
Another signature can be found with the eigenvectors by computing the inverse participation ratio (IPR)  
(see Appendix \ref{sec:overview} for the definition), usually used to quantify 
localization of quantum states \cite{Wegner}. 
For a regular eigenvalue with a well defined continuum eigenvector, the IPR$\,\to 0$ as $1/N_d$ for large $N_d$, while the IPR is much larger within the singular continuum as evidenced by the color code in Fig.~\ref{fig:SpectrumBoxExp}. 
At $a=a^*$, this singular part crosses the regular 
$\lambda_1$ branch, leading to a gap closure.
For $a>a^*$, the singular continuum is dominant and governs relaxation.
In Fig.~\ref{fig:SpectrumBoxExp}, $a^*$ is shown by an arrow. Furthermore here, the structure of the spectrum ensures that $a^*=a_{\text{opt}}$, 
see Fig.~\ref{fig:SpectrumBoxExp};
at this point, $\Lambda(a)$ features a cusp. 
Quite remarkably, the acceptance probability $1-R_n$ at $a=a^*=a_{\text{opt}}$
tends at long times towards $0.455$, 
close to the 50\% rule of thumb alluded to above.

%In the gapless case, $a_{\text{opt}}=a^*$: the crossing of the 
%$\lambda_1$ branch, that decreases when increasing $a$, with the singular continuum,
%the top of which increases with $a$, implies that $a_{\text{opt}}$
%coincides with the localization threshold $a^*$. 

%\begin{figure}[htb]
%  \includegraphics[width=0.5\textwidth]{fig2IPRX2vsaForNlambda1.pdf} 
  %&     \includegraphics[width=0.5\textwidth]{IPRX2vsaForNlambda2.pdf}
%    \caption{Inverse Participation Ratio as a function of jump amplitude $a$, for the same sampling problem as in Fig \ref{fig:SpectrumBoxExp}-bottom (gapless case). The localization transition is signalled by the sharp jump of the IPR at $a\simeq 3.xxx$. This threshold does not depend on number of sites $N_d$, as long as $N_d$ is large enough.
%    {\tt can we change notation from $N$ to $N_d$ to avoid confusion with $\cal N$?}}
%    \label{fig:IPR}
%\end{figure}

\begin{figure}[htb]
    \centering
\includegraphics[width=0.5\textwidth]{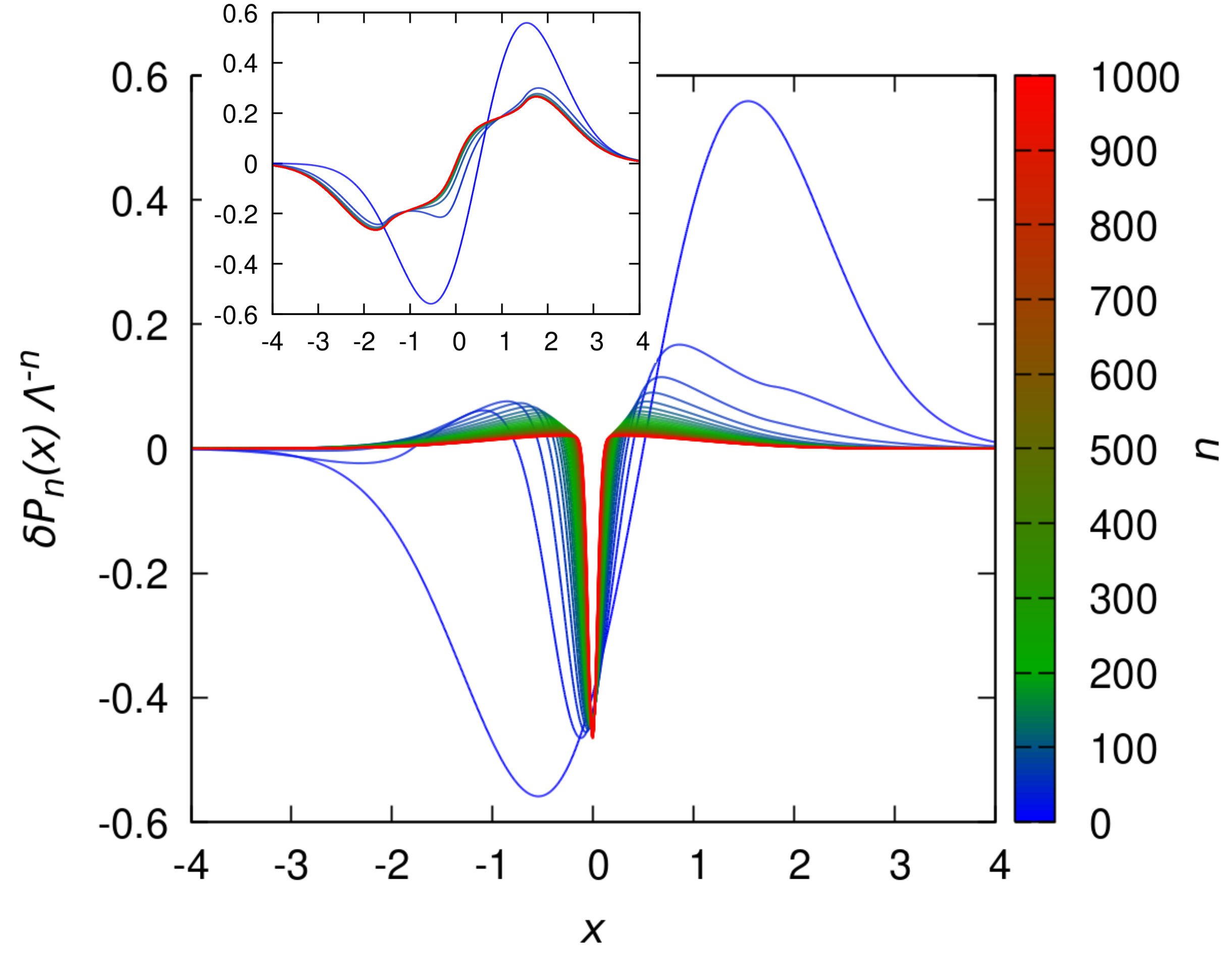}
    \caption{Scaled evolution of the error $\delta P_n(x) = P_n(x) - P_\infty(x)$ vs $x$ for different times $n$
    (indicated by the color code on the right), 
    for the same system as in Fig.~\ref{fig:SpectrumBoxExp}, with the same choice of length unit. Initial $P_0(x) = (2 \pi)^{-1/2} \exp(-(x-1)^2/2)$. 
    Comparison between $a = 3.6>a^*$ (main graph) and $a = 3<a^*$ (inset).
    Although the values of $a$ and the convergence rates are similar in the two graphs, the asymptotic errors are significantly different. }
    \label{fig:relaxX2}
\end{figure}

The critical nature of the parameter $a=a^*$ can be appreciated by 
the behavior of the IPR of the slowest decay mode, as displayed in Fig.~\ref{fig:SpectrumBoxExp}-bottom.
The large value of the IPR for $a>a^*$ indicates that $\delta P_n$
ceases to be spread over the whole system, but rather 
gets more and more ``pinned'' onto a discrete set of points; in the present case,
this set reduces to a single point, $x_l=0$. 
This results in the central dip in the error $\delta P_n(x)$ observed in the main graph
in Fig.~\ref{fig:relaxX2}, that becomes more narrow as time $n$
increases (see below). Fig.~\ref{fig:relaxX2} also reveals that 
a complete change of symmetry goes with the crossing of $a^*$. For $a<a^*$, the longest lived
perturbation in the system is antisymmetric, see the inset of Fig.~\ref{fig:relaxX2}: given the symmetry
of the confining potentials considered ($U(x)=U(-x)$), such a mode takes indeed longer to relax than symmetric ones. 
This can be understood from the mapping of our problem 
to a Schr\"odinger equation, for small $a$, see Appendix \ref{sec:schroedinger}: the first excited state, meaning the $\lambda_1$ branch, has only one zero and is anti-symmetric.
On the other hand, for $a>a^*$, $\delta P_n$ becomes symmetric after a 
transient (see the evolution from an early asymmetric situation
towards symmetry in Fig.~\ref{fig:relaxX2}).

\begin{figure}[htb]
    \centering
\includegraphics[width=0.5\textwidth]{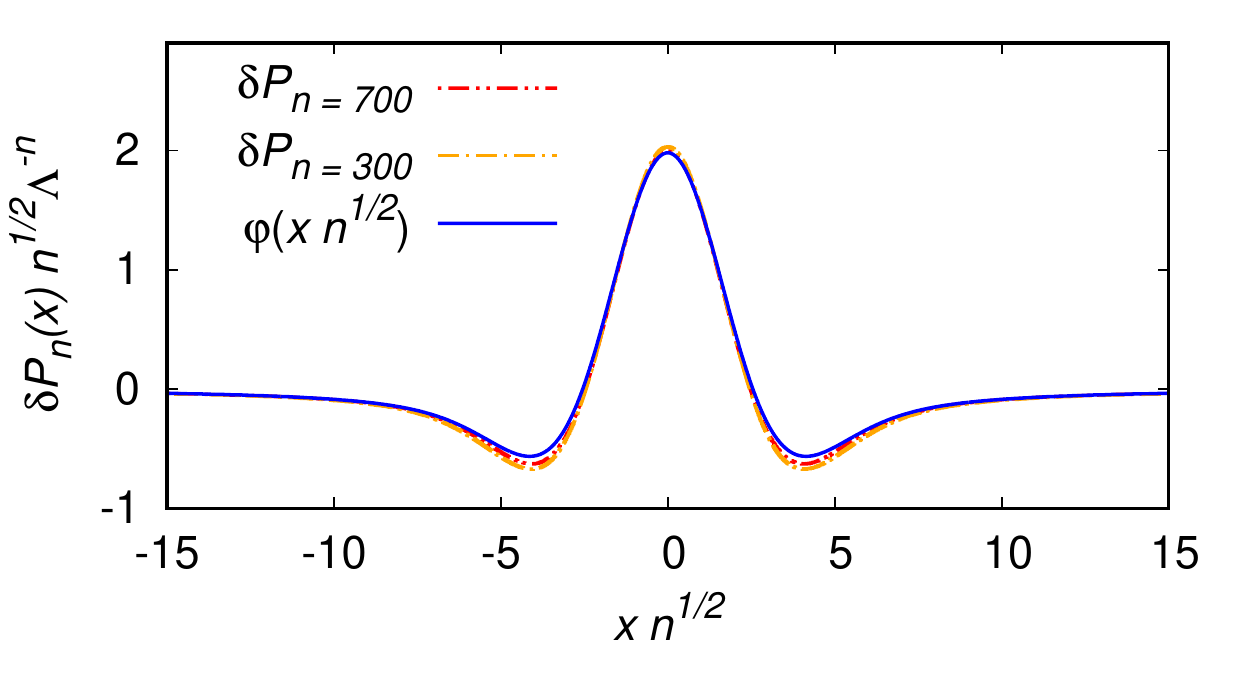}
    \caption{Comparison between the exact calculation presented in Appendix \ref{sec:box} and the numerical data for the scaling behavior of localization. Box potential confinement with $w(\eta) = 3(1+a^{-2}\eta^2)\theta(a-|\eta|)/(8 a)$.
    Lengths are expressed in unit of the box size $L$, convergence to the scaling function is shown for $a = 2.1 $ and $P_0(x) = 2 \theta(1/2-|x|)$. Our analytical expression for the scaling function $\varphi(z)$ and the proof that ${\cal N} = 0$, for this choice of $w(\eta)$, are obtained for $a > 2 > a^* \simeq 1.79$.}
    \label{fig:rescaled}
\end{figure}

To gain more insights into the localization phenomenon and its dynamics, we studied analytically the Master equation 
for confinement in a box, i.e. when $U(x)=0$ for $|x|<L$
and $U(x)=\infty$ for $|x|>L$. Such a case is rich enough to
display the generic phenomenology of localization, while
remaining sufficiently simple  to allow for the derivation of exact results
for several jump distributions $w(\eta)$, see Appendix \ref{sec:box}. 
For cases where $w(\eta)$ is minimum at $\eta=0$, we proved that ${\cal N} = 0$ for sufficiently large $a$ as in the case of harmonic confinement. As in Fig.~\ref{fig:relaxX2}, the localization transition then manifests as a progressive collapse of the error $\delta P_n = P_n(x) - P_\infty(x)$ onto the point where rejection probability is maximal ($x_l=0$), with a spread which decays as $1/\sqrt{n}$.
More precisely, in the vicinity of this point, we obtained the asymptotic form
\begin{equation}
\delta P_n(x) \,=\, \Lambda^n n^{-\gamma} \, \varphi(x\sqrt{n})
\label{eqScaling}
\end{equation}
where $\varphi(z)$ is a regular scaling function, and the exponent 
$\gamma$ depends on $w(\eta)$ and $U(x)$. We found that a scaling function ansatz with $\Lambda = 1$ also describes the relaxation of a zero temperature Metropolis Monte-Carlo algorithms towards a minimum \cite{MCzeroTemp}. In the zero temperature limit, one expects a Dirac delta-function at the 
minimum of the potential. Indeed we found that this expectation is fulfilled. At finite temperature however, the steady state has a finite width which is given by the thermal length, and thus the scaling-function ansatz does not directly follow from the ground state. The difference between the two cases can also be seen from the vanishing integral $\int \varphi(x) dx = 0$ in \eqref{eqScaling} while this integral is normalized to unity at zero temperature. 
Figure \ref{fig:rescaled} shows that 
such a form is well obeyed in the simulations, and that $\gamma=1/2$ for the case
displayed, in full agreement with our exact treatment that 
also explicitly provides $\varphi(z)$  in Appendix \ref{sec:box} (Eq. \eqref{eqScalingPhi}), shown by the continuous line. Numerical evidence shows that for the harmonic potential,
$\gamma=0$. Analytical studies of the box potential where $w(\eta)$ is maximum at $\eta=0$, provide examples where we can prove that ${\cal N} = 1$, in the large $a$ limit. 
The localisation transition in the error can
consequently not be seen for a generic observable, but special choices of the observable or initial conditions allow to reveal a hidden localization transition, even in this case.

\section{Analytical approximation to relaxation rates from Fokker-Planck eigenvectors} 
\label{sec:4_analyt}

We already mentioned that the critical amplitude $a^*$ separates two regimes, a regime $a < a^*$ where the dynamics is governed by the relaxation of diffusion eigenmodes and a regime $a > a^*$ where the relaxation is governed by the highest rejection probability. Surprisingly this knowledge provides a very precise approximation scheme to find quantitatively the full dependence of the relaxation rate $\Lambda(a)$ on the jump length $a$. In the limit of small $a$, the Master equation reduces to a Fokker-Planck equation, and it is possible to use the lowest eigenmodes of this Fokker-Planck equation to project the full Master equation on a small finite dimensional-basis; the details of this procedure are described in  
Appendix \ref{sec:analytical} and illustrated in Fig 
\ref{fig:theoapprox}. We find that for $a < a^*$, a very small number of diffusion eigenmodes provide a very accurate estimation of $\Lambda(a)$ or good analytical approximations when the diagonalization of the reduced matrix is possible. On the contrary, for $a > a^*$, the convergence of this procedure is very slow, and $\Lambda(a)$ coincides with the maximum rejection probability. We notice that the fast convergence of the Fokker-Planck eigenvector expansion was reported previously in \cite{Talbot}, but it was not realized that this fast convergence is limited to the diffusive phase only. 
For a flat jump distribution $w(\eta)$ in a harmonic potential,
for which $a_{opt} = a^*$, this procedure also provides an analytical estimate of the optimal mean acceptance probability, $1-R_\infty \simeq 0.455$, close to values obtained by numerical diagonalization.
A similar computation can be done for Gaussian jumps
(see Appendix \ref{sec:analytical}),
for which we get analytically $1-R_\infty \simeq 0.467$, 
which improves the previously reported numerical estimate
of $0.44$ \cite{Gelman}, alluded to above.

\begin{figure}[htb]
    \centering
\includegraphics[width=0.5\textwidth]{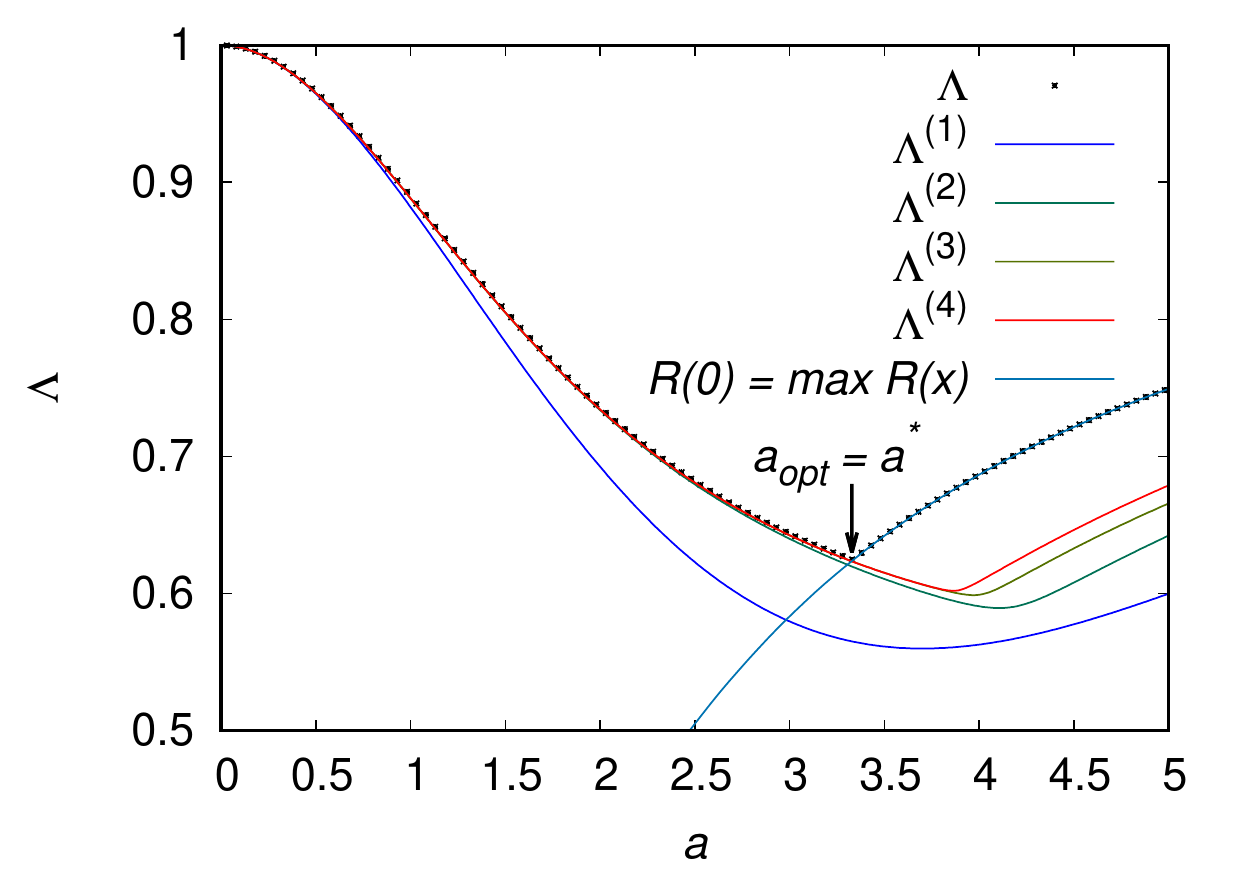}
    \caption{Harmonic confinement with flat jump distribution. Plot of the largest non-stationary eigenvalue as function of jump amplitude $a$, obtained by 1)  exact numerical diagonalization (dots) and 2) analytical approximation in the truncated Schr\"odinger equation basis, with increasing $N_s$, the number of anti-symmetric diffusion modes retained (see Appendix \ref{sec:analytical}). With $N_s=2$, the analytical predictions are already very accurate for $a < a^*$, convergence is very slow on the other side of the localization transition $a > a^*$}
    \label{fig:theoapprox}
\end{figure}

An interesting issue is to assess how robust is the localization transition found: does it survive in higher dimensions 
or in the presence of interactions between particles?
To investigate these, we have studied a) a non interacting model in dimensions 2 and 3, and b), an interacting system 
in dimension 1 and c) the situation where the confining potential features multiple local minima, see Appendix \ref{sec:generalizations}. In all cases, we found a localization transition, demonstrating its wider applicability. 
Analyzing the fate of the present localization transition
for more complex potential landscapes, as found in
disordered systems,
is an interesting open problem.

\section{Conclusions}
\label{sec:5_concl}

To summarize, we have uncovered that a localization transition does generically take
place in Monte Carlo sampling, for a critical
value $a^*$ of the amplitude of the jump distribution.
%for algorithmic parameters
%optimizing the convergence rate, here the amplitude $a$.
%:the transition point coincides with the optimal value,
%$a^*=a_{\text{opt}}$.
%For a bounding potential, the localisation transition is always observed. For the case of the box potential ($U(x) = 0$, except outside a given interval where $U$ diverges), we found examples, depending on the sampling scheme,  where one more delocalized eigenmode can remain, in addition to the equilibrium mode. 
A central result of this paper is 
to show that at $a=a^*$, a singular continuum takes over the regular spectrum as
the leading relaxation mode. We found that below $a < a^*$, the relaxation rate can be determined very accurately by the projection of the full Master equation on the leading relaxation modes of the Fokker-Planck dynamics. This opens the way to analytic calculation for the relaxation rate. For $a > a^*$, the convergence of this expansion becomes much slower and the relaxation rate is instead given by maximal rejection probability.
%indeed the observation that the optimal jump amplitude $a_{\text{opt}}$ coincides with $a^*$ where 
%a singular continuum takes over the regular spectrum as the leading relaxation mode. 
This results in a dynamical collapse,
evidenced by a sharp increase of the  IPR (inverse participation ratio) at $a=a^*$, reminiscent 
of Anderson localization \cite{Anderson_book}. However the 
underlying physical pictures differ. In the Anderson  scenario, the localization length is given by the mean free path of a disordered potential. Here, the error progressively shrinks to a point
with increasing time $n$, without any corresponding limiting eigenvector $\psi(x)$ with non zero norm $\int |\psi(x)|^2 dx > 0$. Thus our study shows an example of a well known Markov process whose relaxation is not determined by the contribution of discrete eigenmodes, but by a progressive localization (collapse) on discrete points.
Furthermore, we found that $a^*$, when it exists,  
coincides with the optimal jump length $a_{opt}$, although we are not able to prove it.
%This remains an interesting open problem.
We may surmise
that the localization phenomenon has been overlooked so far for the reason that 
the upper part of the spectrum, $\Lambda(a)$, which rules relaxation,
is continuous for all $a$ including the transition point $a^*$; it is the derivative 
$d\Lambda/da$ that is discontinuous at $a^*$.
%the Monte Carlo practitioner usually does not search for optimal sampling before
%starting a simulation. 
Yet, the error incurred, due to unavoidable lack of
convergence at finite time, does change nature when crossing $a^*$:
its symmetry, amplitude, and scaling are deeply affected.
The understanding of the localization transition in Monte Carlo
relaxation modes may help to avoid excess events on the localization
sites in the applications of Monte Carlo random walks.

Acknowledgements: We thank M. Rousset (INRIA Rennes) for discussions on the Mathematical aspects of Metropolis algorithm relaxation analysis.

%%%%%%%%%%%%%%%%%%%%%%%%%%%%%%%%%%%%%%%%%%%%%%%%%%%%%%%%%%%
%APPENDICES
\begin{appendix}

\section{Why an optimal jump amplitude?}
\label{sec:optimal}

The relaxation time of the Metropolis algorithm, for a given functional form $w$ of the jumps (see Eq.~\eqref{eq:w_f} below),
depends  on the jump amplitude $a$. On general grounds,
this time should exhibit 
a non-monotonous behavior with a well-defined minimum
at some specific amplitude $a_{\text{opt}}$
(corresponding to a minimum convergence time $\tau$ (minimum $\Lambda$, i.e. a maximum rate $-\ln\Lambda = 1/\tau$). This is the so-called Goldilock’s principle
\cite{Goldilocks}.
The rationale behind this expectation goes as follows:
\begin{itemize}
\item In the diffusive limit where $a$ is small, though most of the
jumps are accepted, the particle moves over a limited region of 
space which results in a long time for
exploring the full available space. Hence, we expect $\tau$
to diverge, i.e. $\Lambda \to 1$. We can be more specific, assuming
a confinement potential of the form $U(x)=|x|^\alpha$ with $\alpha>0$. 
At equilibrium,
the walker's density $P$ will be concentrated within the thermal length 
$\ell \propto \beta^{-1/\alpha}$ around the origin, and equilibrium will be 
reached after a characteristic time $\tau$ such that $D\tau= \ell^2$,
where $D$ is the diffusion coefficient.
For our discrete time dynamics, we have $D\propto a^2$, so that we expect
here $\tau \propto \beta^{-2/\alpha} a^{-2}$, meaning 
$\Lambda -1 \propto a^2$
\item In the opposite long jump limit with large $a$,  most of the moves are rejected and the particle hardly moves. As long as $w(0)$ is non-vanishing, increasing the jump amplitude $a$ simply reduces the 
displacement probability by a factor $1/a$, while leading to the 
same sampling of phase space on the scale of the confinement length $\ell \ll a$. 
Hence we expect
the system to relax very slowly, i.e., the relaxation time $\tau$ to diverge
as $\tau \propto a$, so that $\Lambda -1 \propto 1/a$. This scaling law can only be altered for $w(0)=0$. 
\end{itemize}
We thus expect an optimal finite jump amplitude $a=a_{\text{opt}}$, for a given functional form $w$, where $\Lambda(a)$ is minimal and hence the
convergence is the fastest.

%%%%%%%%%%%%%%%%%%%%%%%%%%%%%%%%%%%%%%%%%%%%%%%%%%%%%%%%%%%%%%%%%%%%%%%%
\section{The Master equation and its reformulations}
\label{sec:master}

\subsection{The formalism}

From the dynamics defined in the main text, we can write the Master equation obeyed by the walker's density as 
\begin{equation}
P_n(x) =  \int_{-\infty}^{\infty} dx'\, P_{n-1}(x')\, w(x-x')\, {\rm min}\left(1, e^{-\beta\, \left(U(x)-
U(x')\right)}\right) +
\left[1- \int_{-\infty}^{\infty} dy\,  w(y-x)\, {\rm min}\left(1, e^{-\beta\, \left(U(y)-U(x)\right)}\right)\right]\, P_{n-1}(x)\, , 
\label{Master_T.1}
\end{equation}
where $w(\eta)$ is the jump distribution and $U(x)$ the confining potential. 
At a given time step $n$, the first term describes the probability flux to $x$  from all other positions $x'$. 
The second term is for the probability that all attempted moves made by the particle at $x$ (to
another arbitrary position $y$) are
rejected.
It proves convenient to replace the `min' function above by the identity
\begin{equation}
{\rm min}\left(1, e^{-\beta\, \left(U(x)-U(x')\right)}\right)=
\theta\left(U(x')-U(x)\right) + e^{-\beta\, \left(U(x)-U(x')\right)}\, \theta\left(U(x)-U(x')\right)\, 
\label{min_def}
\end{equation}
where $\theta(z)$ is the Heaviside theta function. 
The Master equation \eqref{Master_T.1} can then be written as
\begin{equation}
P_n(x)= \int_{-\infty}^{\infty} F_\beta(x,x')\, P_{n-1}(x')\, dx'
\label{Master_T.2}
\end{equation}
where the temperature dependent kernel is given by
\begin{eqnarray}
F_\beta(x,x') &= & w(x-x')\, \left[\theta\left(U(x')-U(x)\right)+ e^{-\beta\,
\left(U(x)- U(x')\right)}\, \theta\left(U(x)-U(x')\right)\right] \nonumber \\
& & + \delta(x-x')\,
\underbrace{\left[   1- \int_{-\infty}^{\infty} dy\, w(y-x')\,  \left[ \theta\left(U(x')-U(y)\right)+ e^{-\beta\,
\left(U(y)- U(x')\right)}\, \theta\left(U(y)-U(x')\right)\right]\right]}_{R(x')} \, .
\label{kernel_T.1}
\end{eqnarray}
The kernel $F_\beta(x,x')$ can be interpreted as the probability of a jump from $x'$ to $x$
at inverse temperature $\beta$. The term in square brackets on the second line of Eq.~\eqref{kernel_T.1} is the rejection probability, that can be recast in
\begin{equation}
    R(x')  = \int_{-\infty}^{\infty} dy\, w(y-x') \left(1 - e^{-\beta \, \left(U(y)- U(x')\right)} \right) \, \theta\left(U(y)-U(x')\right) .
\label{eq:Rejection_rate}
\end{equation}
Written as such, it directly expresses the fact that 
among all attempted moves from $x'$ to $y$, only a fraction 
$1-e^{-\beta \, \left(U(y)- U(x')\right)}$ of those leading to an energy increase ($U(y)>U(x)$), is effectively rejected. All others attempts are accepted and thus 
do not contribute to $R(x')$.

A first check for the validity of the Master equation is that
it should conserve the total probability $\int_{-\infty}^{\infty} P_n(x)\, dx=1$.
From \eqref{Master_T.2}, this means that kernel $F_\beta(x,x')$ must satisfy the 
condition
\begin{equation}
\int_{-\infty}^{\infty} F_\beta(x,x')\, dx=1 \quad {\rm for} \,\, {\rm all}\,\, x' \, .
\label{sum_rule_T.1}
\end{equation}
Indeed, substituting $F_\beta(x,x')$ from \eqref{kernel_T.1} into the integral
\eqref{sum_rule_T.1}, it is easy to check that it satisfies the probability conservation for all
$x'$.  

Next, we verify explicitly that the Master equation \eqref{Master_T.2}, with $F_\beta(x,x')$
given in \eqref{kernel_T.1}, admits, as $n\to \infty$, a stationary solution that is of the 
Gibbs-Boltzmann equilibrium form 
\begin{equation}
    P_\infty(x) \,=\, \frac{1}{Z} e^{-\beta U(x)},
    \label{eq:GibbsBoltzmann}
\end{equation} 
where the partition function $Z$ is a normalization constant. 
Assuming a stationary solution exists as $n\to \infty$
in \eqref{Master_T.2}, it must satisfy the integral equation
\begin{equation}
P_{\infty}(x)= \int_{-\infty}^{\infty} F_\beta(x,x')\, P_{\infty}(x')\, dx' \, .
\label{Pstat_T.2}
\end{equation}
To verify this equality, we substitute $ P_{\infty}(x')= (1/Z)\, e^{-\beta\, U(x')}$ on
the right hand side (rhs) of \eqref{Pstat_T.2} and use the explicit form of $F_\beta(x,x')$ from
\eqref{kernel_T.1}. By writing down each term on the rhs explicitly, it is straightforward to
check that indeed for arbitrary symmetric jump distributions such that $w(x-x')=w(x'-x)$, the
rhs gives (after a few cancellations) $(1/Z) e^{-\beta\, U(x)}$ for arbitrary confining potential 
$U(x)$. 
This is of course expected since the Metropolis rule indeed
%\eqref{Metropolis_T.1} 
does satisfy 
detailed balance with respect to the Gibbs-Boltzmann stationary state.
    
%%%%%%%%%%%%%%%%%%%%%%%%%%%%%%%%%%%%%%%%%%%%%%%%%%%%%%%%%%%%%%
\subsection{Transformation to a self-adjoint problem}

Solving the Master equation (\ref{Master_T.2}) analytically 
for arbitrary potential is out of reach. A first difficulty
one encounters is that the kernel $F_\beta(x,x')$ in \eqref{kernel_T.1} is non-symmetric
under the exchange of $x$ and $x'$: the integral operator $F_\beta(x,x')$ is not
self-adjoint. This problem can be circumvented by applying the following `symmetrizing' trick \cite{Risken}.
Let us first define a new quantity $Q_n(x)$ related simply to $P_n(x)$ via the relation
\begin{equation}
P_n(x)= e^{-\beta\, U(x)/2}\, Q_n(x) \, .
\label{Pn_Qn.1}
\end{equation}
Substituting this relation in \eqref{Master_T.2}, we see that $Q_n(x)$ satisfies the
following integral equation
\begin{equation}
Q_n(x)= {\widehat K}_\beta Q_{n-1}(x) = \int_{-\infty}^{\infty} K_\beta(x,x')\, Q_{n-1}(x')\, dx'
\label{integral_Q.1}
\end{equation}
where the action of the integral operator ${\widehat K}_\beta$ is described by its kernel $K_\beta(x,x')$:
%\begin{eqnarray}
\begin{align}
K_\beta(x,x') = w(x - x') e^{-\beta | U(x) - U(x') |/2} + \delta(x - x') R(x)    
\label{K_symm_T.1}
\end{align}
and the rejection probability $R(x)$ is defined in Eq.~(\ref{eq:Rejection_rate}).
Thus, for symmetric jump distribution $w(x-x')=w(x'-x)$, $K_\beta(x,x')$
is symmetric and 
we can consider $\widehat K_\beta$ as a real self-adjoint integral operator (operating on the real line) whose
matrix element $\langle y|\widehat K_\beta|y'\rangle= K_\beta(y,y')$ is given by Eq.~(\ref{K_symm_T.1}). 
Besides, Eq.~(\ref{integral_Q.1}) admits a stationary solution
\begin{equation}
Q_{\infty}(x)= \frac{1}{Z}\, e^{-\beta\, U(x)/2} .
\label{Q_stat.1}
\end{equation}

The solution of the
integral equation (\ref{integral_Q.1}) can be written as a linear combination of the eigenmodes
of the operator $\widehat K_\beta$, i.e.,
\begin{equation}
Q_n(x)= \sum _{\lambda} {\cal A}_\lambda\, \psi_{\lambda}(x)\, \lambda^n
\label{eq:phi_lambda}
\end{equation}
where $\psi_{\lambda}(x)$ satisfies the eigenvalue equation
\begin{equation}
\int_{-\infty}^{\infty} K_\beta(x,x')\, \psi_{\lambda}(x')\, dx'= \lambda\, \psi_{\lambda}(x)\, 
\label{eq:psi_lambda_eigenvalue}
\end{equation}
and the ${\cal A}_\lambda$'s are arbitrary at this point. Consequently, from Eq.~\eqref{Pn_Qn.1},
\begin{equation}
    P_n(x) \,=\, \sum_\lambda {\cal A}_\lambda \,   \psi_\lambda(x) \,e^{-\beta U(x)/2} \lambda^n \,=\,   \sum_\lambda  {\cal A}_\lambda \, {\cal P}_\lambda(x)  \lambda^n \quad \hbox{with} \quad 
    {\cal P}_\lambda(x) \,=\,  \psi_\lambda(x) \,e^{-\beta U(x)/2} ,
\end{equation}
as written in Eq. (9) in the main text.

Since the operator $\widehat K_\beta$ is real self-adjoint, both its eigenvalues and eigenvectors are real valued \cite{Levin}. 
This property extends to the operator defined from $F_\beta(x,x')$, since
\begin{equation}
    e^{-\beta U(x)/2} \,F_\beta(x,x') \,=\, e^{-\beta U(x')/2} \,K_\beta(x,x').
\end{equation}
Having a real spectrum is a non-trivial property, as the eigenvalues of Frobenius-Perron type of operators to which the original integral equation Eq~(\ref{Pstat_T.2}) belongs are in general complex numbers inside the unit circle $|\lambda|<1$. The detailed balance rules which are used to derive the Metropolis algorithm actually constrain the eigenvalue of the associated integral equation to be real (at non zero temperatures) \cite{Levin}. 
The eigenvalue $\lambda_0=1$ corresponds to the steady state solution $Q_{\infty}(x)$ in \eqref{Q_stat.1};
all other eigenvalues are real and strictly below 1.
We have labeled the spectrum so that  $1>\lambda_1 \geq \lambda_2\ldots$. 
A particular interest goes into the eigenvalue $\lambda_1$ that is closest to $1$ from below, since it rules the long time dynamics.

%%%%%%%%%%%%%%%%%%%%%%%%%%%%%%%%%%%%%%%%%%%%%%%%%%%%%%%%%%%%%%
\subsection{The diffusive limit: Schr\"odinger reformulation and symmetry}
\label{sec:schroedinger}

The distribution $w$ of attempted jumps is taken of the form \eqref{eq:w_f}
%\begin{equation}
%    w(\eta) \,=\, \frac{1}{a} f\left(\frac{\eta}{a}\right) ,
%    \label{eq:w_f}
%\end{equation}
with $a$ representing a characteristic length. 
The limit of small $a$ is informative: 
the original Master equation reduces to a diffusive-like
Fokker-Planck equation \cite{Risken}. 
In line with our preceding treatment, it is more convenient
to work with the self-adjoint dynamics, which is described by an equivalent Schr\"odinger equation, as we proceed to show. 

For $a\to 0$, it is possible to Taylor-expand the eigenvalue equation 
\eqref{eq:psi_lambda_eigenvalue}. Introducing the second moment of the jump distribution 
\begin{align}
\sigma^2 &= \int dy \; y^2 w(y) ,
\label{eqDEFY2}
\end{align}
and making use of the identity 
 $\int dy \; w(x - y) (x - y) = 0$ together with the symmetry 
 relations, we get
\begin{align}
 \int dy \; w(x - y) \frac{(x-y)^2}{2} {\rm sign}( U(x) - U(y) ) =  \int dy \; w(x - y) \frac{(x-y)^2}{2} {\rm sign}( U'(x) (x -y) ) = 0
\end{align}
valid when $U'(x) \ne 0$ and in the limit of small jumps. These cancellations stem from the 
symmetry $w(x)=w(-x)$.
We thereby get:
\begin{align}
(1 - \lambda) \psi_\lambda(x) &= \sigma^2 \left( -\frac{1}{2} \psi_\lambda''(x)  + \frac{\beta^2}{8}  U'(x)^2 \psi_\lambda(x) -   \frac{\beta}{4}  U''(x) \psi_\lambda(x) \right)
\label{eqSmallJump}
\end{align}
The relaxation rates of this equation can thus be determined from the eigenvalues $\epsilon_n$ and eigenvectors of the effective Schr\"odinger equation
\begin{align}
{\widehat H} \psi = -\frac{1}{2} \psi''(x)  + \frac{\beta^2}{8}  U'(x)^2 \psi(x) -   \frac{\beta}{4}  U''(x) \psi(x) = \epsilon_n \psi(x) .
\label{eqH}
\end{align}
The connection reads 
\begin{equation}
    \lambda_n \, =\, 1 - \sigma^2 \epsilon_n,
    \label{eq:lambda_energy_Schro}
\end{equation}
providing an explicit expression for the spectrum 
$\{\lambda_n\}$.

We stress that a truncation of the Taylor expansion
behind the derivation of the Schr\"odinger equation is justified if the length-scale on which the wavefunctions vary is large compared to $a$, the typical amplitude of the jumps generated by $w(\eta)$. Thus Eq.~(\ref{eqSmallJump}) is not valid in the limit of the high energy modes $\epsilon_n$ of the Schr\"odinger equation. This limitation of the Schr\"odinger picture can be anticipated from the fact that the eigenvalues of the original Master equation are in the interval $\lambda \in [-1,1]$ while the eigenvalues predicted by Eq.~(\ref{eqSmallJump}) extend to all the range $(-\infty, 1]$.
The ground state of the Hamiltonian Eq.~(\ref{eqH}) has a vanishing ground state eigenvalue $\epsilon_0 = 0$ with an eigenvector given by $\psi_0 = e^{-\beta U(x)/2}$. This eigenvector describes the equilibrium probability distribution and is identical to the ground state of the original Eqs.~(\ref{integral_Q.1},\ref{K_symm_T.1}), without the assumption of a small jump length.

Note that since the original confining potential is symmetric in $x$ (even),
so is the effective potential in the  Schr\"odinger Eq.~\eqref{eqH}, 
$ \beta^2  U'(x)^2 /8 -  \beta  U''(x)/4$.
The Schr\"odinger reformulation then allows to understand why the longest lived eigenmode,
for small $a$, is antisymmetric: it corresponds to the first excited state,
with an eigenfunction featuring a unique zero. %Besides, the  Schr\"odinger representation is at the root of analytical progress, see below.

%%%%%%%%%%%%%%%%%%%%%%%%%%%%%%%%%%%%%%%%%%%%%
\subsection{Analytical solutions in the truncated Schr\"odinger eigenbasis} 
\label{sec:truncated_schroedinger}

Since Eq.~(\ref{eqH}) is a  Schr\"odinger equation, its (normalized) excited state  eigenvectors $\psi_n(x) \;(n \ge 1)$ are all orthogonal to $\psi_0(x)$ and provide a natural basis for a variational estimation of the relaxation rate.
Indeed, the definition of $\Lambda$ in the main text as the 
leading relaxation mode (upper value of the relaxation spectrum, leaving aside the top eigenvalue $\lambda=1$
corresponding to the equilibrium state) can be recast 
as
\begin{equation}
    \Lambda \, =\, \max_{\Phi \perp \psi_0} 
    \frac{\int dy\, \int dy'\,  \Phi(y) K_\beta(y,y') \Phi(y')}{\int \Phi^2(y)\, dy} .
\end{equation}
As a consequence, by restricting to the
 first $N$ excited states (which are perpendicular to the ground state $\psi_0(x)$), we get a lower bound in the form
% A convenient approximation (and lower bound) for $\lambda_1$ can thus be obtained by selecting the first $N$ excited state eigenvectors and evaluating: 
\begin{align}
\Lambda &\ge \max_{c_1,...c_{N_s}} \frac{\int dy\, \int dy'\,  \Phi(y) K_\beta(y,y') \Phi(y')}{\int \Phi^2(y)\, dy} \\
&\Phi = c_1 \psi_1 + ... + c_{N_s} \psi_{N_s}
\label{eqSmBasis}
\end{align}
When the Schr\"odinger equation limit is valid, Eqs.~(\ref{eqSmallJump})-\eqref{eq:lambda_energy_Schro}
allow to approximate the relaxation rates of the Metropolis algorithm from the eigenvalues of the Schr\"odinger equation:
\begin{align}
    \lambda_n = 1 - \sigma^2 \epsilon_n .
\end{align}
Upon increasing of the typical size of jump length, the operator $\widehat K_\beta$ will mix  different Schr\"odinger eigenmodes and this estimate will no longer be valid.

Solving the present optimization problem is equivalent to finding the largest eigenvalue of the reduced 
$N_s\times N_s$ matrices $K^{(N_s)}$ with matrix elements
\begin{align}
K_{nm} = \int dy\, \int dy'\,  \psi_n(y) K_\beta(y,y') \psi_m(y'),
\label{eqKNs}
\end{align}
with the truncation $1 \leq n, m \leq N_s$ where the positive integer $N_s$ gives the number of retained eigenfunctions.
We will show in section \ref{sec:analytical} that with  a few modes only,
very good quantitative estimates for $\Lambda$ can be obtained by this approach, even where Eq.~(\ref{eqSmallJump}) is no longer valid, far from the small jump amplitude limit. 

In cases where the potential $U(x)$ is even (as assumed here), the eigenbasis $\psi_n(x)$ will split into symmetric and anti-symmetric eigenfunctions. The Master equation kernel $K_\beta$ inherits the symmetry properties of the potential $U(x)$ and the matrix elements Eq.~(\ref{eqKNs}) will be non-zero only for wavefunctions from the same parity. The truncated matrix will thus split into a direct sum of even-even and odd-odd matrices. The mapping to the Schr\"odinger equation ensures that at least in the small jump limit, $\Lambda$ will be in the odd sector, but we will show
in section \ref{sec:analytical}
an example where this is not necessarily true for large $a$.

%%%%%%%%%%%%%%%%%%%%%%%%%%%%%%%%%%%%%%%%%%%%%%%%%%%%%%%%%%%%%%%%%%%%%%%%%%
\section{Overview of the cases investigated and main tools of analysis}
\label{sec:overview}

\subsection{Potentials, sampling choice, and observables}
\label{ssec:observables}

The claims put forward in the main text rely on the study of a number 
of confining potentials of the form $U(x) \propto |x|^\alpha$,
with $\alpha>0$. Some emphasis has also been put in the study of confinement by hard
walls, the box potential,  where $U(x)=0$ for $x\in [-L,L]$ and $U(x)=\infty$ for $|x|>L$. 

In these potential landscapes, we have changed the sampling method,
varying the distribution $f(\eta)$ of attempted jumps. Scaling out the 
jump's typical length $a$, we obtain the dimensionless distribution 
$f(z)$:
\begin{equation}
    w(\eta) \,=\, \frac{1}{a} f\left(\frac{\eta}{a}\right) .
\end{equation}
Different choices were made, symmetric for simplicity ($f(z)=f(-z)$):
\begin{itemize}
\item Gaussian distribution of jumps
\begin{equation}
    f(z) \, = \, \frac{1}{\sqrt{2\pi}} \,e^{-z^2/2}
\end{equation}
\item Exponential distribution
\begin{equation}
    f(z) \, = \, \frac{1}{2} e^{-|z|}
\end{equation}
\item Flat distribution
\begin{equation}
    f(z) \, = \, \theta\left(\frac{1}{2}-|z|\right)
    \label{eq:flatw}
\end{equation}
\item Other more specific choices, as introduced to analyze the
box confinement, see section \ref{sec:box}.
\end{itemize}

In order to study convergence to equilibrium, it is important to 
pay attention to the symmetry of the observables used,
for it affects relaxation rates. This can be understood
from the Schr\"odinger reformulation,
where excited states of increasing order are alternatively 
even and odd in $x$, while their energy is directly related to
the relaxation rate, see Eq.~\eqref{eq:lambda_energy_Schro}. 
%Therefore, by a proper choice of observables, one can distinguish...
Therefore, we can use even observables (with even initial conditions) to suppress
a slower relaxation rate corresponding to an odd mode, allowing to estimate $\lambda_1$
and $\lambda_2$ from sampling. In particular, we measured 
\begin{equation}
\mathcal{O}_1(x) = (x - 0.5)^2, \quad \hbox{ and } \quad
\mathcal{O}_2(x) = |x| .
%,\quad \mathcal{O}_3(x) = |x + 0.5|, \quad
%\mathcal{O}_4(x) = \theta(1/32 - |x|)  .  
\label{eq:observables}
\end{equation}
% I put here all I measured and comment all which are not used
%$\mathcal{O}_1(x) = |x|$,
%$\mathcal{O}_2(x) = |x + 0.5|$,
%$\mathcal{O}_3(x) = (x - 0.5)^2$, and
% $\mathcal{O}(x) = x^2$,
% $\mathcal{O}(x) = x^4$,
% $\mathcal{O}(x) = x$,
% $\mathcal{O}(x) = 1 / |x|$,
%$\mathcal{O}_4(x) = \theta(1/32 - |x|)$. %, for $z = 1, 2, 4, 8, 16, 32, 64, 128, 256, 512$
%We use multiple asymmetric observables, since we found that depending on the potential, the
%jump distribution and the initial condition, we could get better quality estimates with 
%different observables. \hen{not true anymore}

%%%%%%%%%%%%%%%%%%%%%%%%%%%%%% 
\subsection{Probing localization with the Inverse  Participation Ratio}

Since the transition we identify amounts to a localization of 
the convergence error onto well defined positions, it is essential 
to discriminate delocalized states, from localized ones. To this end,
we discretize the integral in the Master equation Eq.~\eqref{Master_T.1}
into a sum of $N_d$ terms, with a running position index to
denote lattice sites $1 \leq i  \leq N_d$.
We then introduce the inverse participation ratio for an eigenvector
$\Psi_\lambda(x)$ as
\begin{align}
{\rm IPR}(\lambda) =  \left. \sum_{i=1}^{N_d} |\Psi_\lambda(i)|^4 \middle/ \left( \sum_i |\Psi_\lambda(i)|^2 \right)^{2}\right. .
\label{eqDefIPR}
\end{align}
This quantity can vary between two extremes. 
If the eigenfunction is completely delocalized over the whole system,
so that $\Psi_\lambda(i)$ is a constant (normalization is irrelevant here),
then $\text{IPR}(\lambda)=1/N_d$, with $N_d \gg 1$. If on the other hand,
$\Psi_\lambda(i)$ vanishes on all sites but one, then $\text{IPR}(\lambda)=1$,
irrespective of $N_d$. If $\Psi_\lambda(i)$, the discretization of an eigenvector $\psi_\lambda(x)$, is well defined in the continuum limit, then $\text{IPR} \rightarrow 0$ as $N_d^{-1}$. On the other hand, if part of the eigenfunction localizes, a slower decay
as a function of $N_d$ will be observed and the discrete eigenfunction $\Psi_\lambda(i)$ will not converge to a well defined continuum limit.

%%%%%%%%%%%%%%%%%%%
\subsection{Numerical diagonalization}

Numerical diagonalization of the discretized form of the Master equation Eq.~\eqref{Master_T.1} allows to find the spectrum of eigenvalues. The Master equation was discretized by a uniform mesh with $N_d$ sites. The integration was replaced by a sum over accessible neighbors, ensuring probability conservation. For particles in a box $x \in [-L,L]$, the first and last points of the mesh were set to $-L$ and $L$ respectively. For the harmonic potential, the first and last points were set to $\pm X_{\text{max}}$ where $X_{\text{max}}$ is the largest $|x|$ allowed by the mesh. The results in Fig.~1 from the main text were obtained for $X_{\text{max}} = 10$ (in units of thermal length in the harmonic potential). We increased $X_{\text{max}}$ up to $30$ to check that the results were independent on this choice of $X_{\text{max}}$. To obtain eigenvalues and eigenvectors, we used the diagonalization routines from the eigen++ library. To avoid the appearance of spurious complex eigenvalues due to rounding errors in diagonalization algorithms, we took advantage of detailed balance to transform the kernel of the integral Master equation into its symmetric form Eq.~(\ref{K_symm_T.1}). Considering the fast increase of the numerical time required for full diagonalization with matrix size, we used this approach for $N_d \le 10^4$.

%%%%%%%%%%%%%%%%%%%
\subsection{Numerical iteration of the Master equation}
To study the relaxation of the error 
$\delta P_n(x)$, it is also possible to follow the evolution of a fixed initial state by successive iterations of the Master equation.  This approach is computationally less demanding than full diagonalization.
%for a large number of discretization sites. 
With this method, we ran simulations up to $N_d= 2\times 10^5$.

%%%%%%%%%%%%%%%%%%%
\subsection{Monte Carlo simulations}

To put our analytical calculations to the test and assess the accuracy 
of the predicted bounds, we have directly simulated
the dynamics defined by the Master equation. The Metropolis rule, 
spelled out in the main text, defines a Markov chain which can be readily simulated by means of classical Monte Carlo. 
For large enough time $n$, equilibrium will be reached and the
walker's position will sample the Gibbs-Boltzmann distribution \eqref{eq:GibbsBoltzmann}.
The sampling scheme obeys detailed balance \cite{FrenkelSmith},
which guarantees the existence of a steady state, that is furthermore
unique for an ergodic irreducible chain \cite{Wasserman}.
We are interested in the long-time approach towards the equilibrium distribution. To gather statistics, we perform the simulation until $n=30$ typically, 
and repeat this for $m=10^{10}$ or $10^{11}$ independent samples.
At every time step $n$, we compute a number of observables, see section
\ref{ssec:observables}. An observable $\mathcal{O}$ is then
averaged over all $m$ samples at fixed time $n$, leading to  
$\overline{\mathcal{O}}$. 
\begin{equation}
   \overline{\mathcal{O}}(n) \,=\, \frac{1}{m} \sum_{i=1}^m {\cal O}^{(i)}(n)
\end{equation}
where the observable measured at time $n$ in the $i$th sample is ${\cal O}^{(i)}(n).$

Our Monte Carlo estimates of the largest eigenvalue are obtained by fits 
to the deviation from the equilibrium value at time $n$ of the form
$|\overline{\mathcal{O}}(n) - \left< \mathcal{O} \right>_\text{eq}| = c_1 \lambda_1^n + c_2 \lambda_2^n$, where $\left< \mathcal{O} \right>_\text{eq}$ is the equilibrium value, reached after long times. 
We exclude the first values (typically $n<5$) to minimize influence of transient behavior and $c_1$ and $c_2$ are free constants. 
Technically, we use an analytical value for $\left< \mathcal{O} \right>_\text{eq}$, if known, or Monte Carlo results at $n=200$, where the statistical error dominates over the systematic deviation.
Performing this procedure for multiple jump distributions
parametrized by $a$ allows us to gather measurements of relaxation rates, which we can compare to 
our analytical results and the other computational approaches.
    %\caption{Comparison between the exact calculation presented in \cite{suppl} and the numerical data for the scaling behavior of localization. Box potential confinement with $w$...
%but the box potential has already been introduced;
%\hen{I am not sure what happened here, I hope nothing got lost...}
For a reliable fit, it is necessary to have good estimates of the standard errors of the measured 
mean values; \emph{Welford's algorithm} has been used \cite[p.~232]{knuth1998art}.
%To calculate the necessary means and variances from the data without the need to save all $10^{11}$
%samples \footnote{corresponding to 800 GB of data}, we use \emph{Welford's algorithm} \cite[p.~232]{knuth1998art}.
The acceptance probability is computed during an independent simulation of a single particle over $1.1\cdot10^6$
Metropolis steps, where the first $10^5$ steps are ignored for the average.

\begin{figure}[htb]
    \centering
    \includegraphics[scale=1]{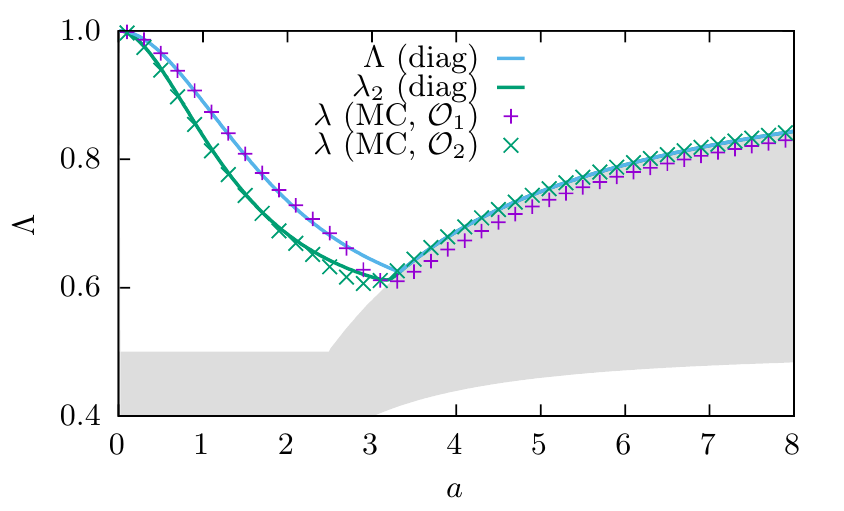}  
    
    \caption{
        %Convergence rate $\Lambda(a)$ for jumps from the distribution $w^\cap$ (left) and $w^\cup$ (right) in a box potential.
        %We compare the results obtained from direct Monte Carlo simulations (symbols), the diagonalization 
        %approach (lines), for simplicity only $\lambda_1$, $\lambda_2$ and the singular spectrum are shown.
        %Both methods agree excellently for both cases of $w(\eta)$.
        %\emma{since the box case is worked out below, we may show only the third graph, and write :} 
        Convergence rate $\Lambda$
        vs. jump amplitude $a$ for a harmonic confinement ($U(x) = x^2/2$)
        and a flat $w(\eta)$ distribution as in Eq.~\eqref{eq:flatw} (cf. Fig.~1 of the main text).
        Comparison between the direct Monte Carlo simulations measure (MC) and the numerical diagonalization technique. Both methods agree very well. The Monte Carlo approach needs to look at symmetric and asymmetric observables to measure the different branches of the spectrum.
        The observables used are provided in Eq.~\eqref{eq:observables}.
        For $a<a^*\simeq 3.33$, using the asymmetric observable 
        ${\cal O}_1$ provides a very good estimate of the largest eigenvalue $\lambda_1=\Lambda$. For $a<a^*$, using the even observable ${\cal O}_2$ yields the second largest eigenvalue
        $\lambda_2$. For $a>a^*$, the largest of both 
        ($\lambda$ from ${\cal O}_2)$ gives a very good estimation 
        of $\Lambda$. The singular continuum is shown by the grey region.
        As in the main text, $a$ is given in units of the thermal length at equilibrium.
    }
    \label{fig:cmp_MC1}
\end{figure}

We verify the quality of the three numerical approaches by comparing them to each other,
and to the analytical results for the case of the box potential. Figure~\ref{fig:cmp_MC1} shows the estimates
of $\Lambda$ obtained from the Monte Carlo simulations and $\lambda_1$ obtained from the 
diagonalization (the upper envelope of the spectrum). 
% for two choices of the jump distribution $w(\eta)$. 
%Both agree excellently with the analytical result, which will be derived in the next section.

\begin{figure}[htb]
    \centering
    \includegraphics[width=0.48\textwidth]{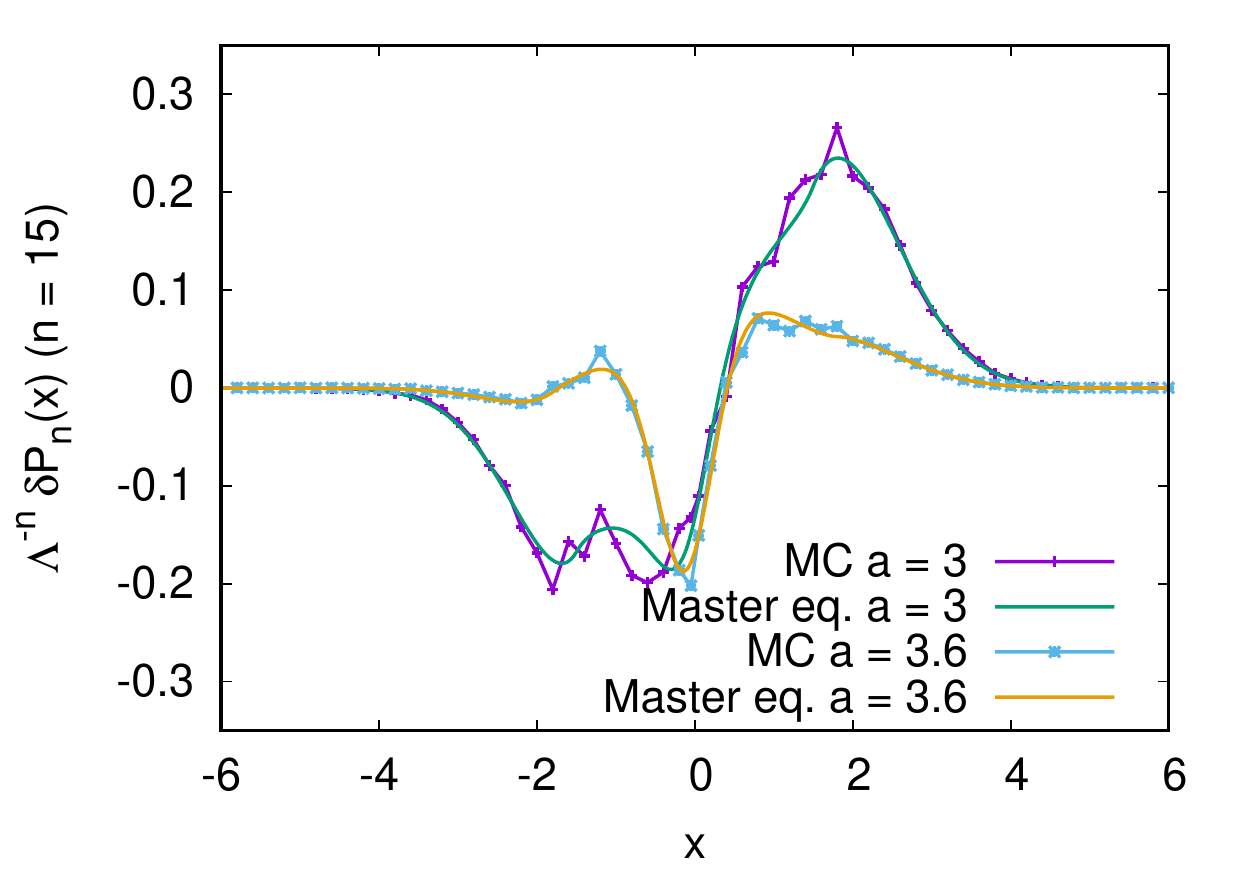}
    \caption{
        Shape of the rescaled deviation $\delta P_n(x)$ vs $x$ from the equilibrium distribution
        at finite times (here $n=15$). The symbols show results of direct Monte Carlo
        and the lines show the results of the iteration of the Master equation; both methods  
        are in good agreement. Same confinement and jump distribution as in Fig.~\ref{fig:cmp_MC1}.
    }
    \label{fig:cmp_MC2}
\end{figure}

Figure~\ref{fig:cmp_MC2} shows the shape of the deviation $\delta P_n(x)$ from 
the equilibrium distribution at a finite time, for uniform jumps in the harmonic potential.
%Since the deviation shrinks exponentially in $n$, it falls below the statistical 
%error of our Monte Carlo simulations very fast, such that direct Monte Carlo is 
%limited to rather small values of $n$.
The direct Monte Carlo
results and the results from the iteration of the Master equation are compatible within statistical 
fluctuations. This lends a high confidence in the results of the iteration
for longer times, which are shown in the main manuscript.

%%%%%%%%%%%%%%%%%%%%%%%%%%%%%%%%%%%%%%%%%%%%%%%%%%%%%%%%%%%%%%%%%%%%%%%%%%

%%%%%%%%%%%%%%%%%%%%%%%%%%%%%%%%%%%%%%%%%%%%%%%%%%%%%%%%%%%%%%%%%%%%%%%%%%
\section{The box potential}
\label{sec:box}

Obtaining exact analytical results in the general case of a potential $U(x)$ in 
$|x|^\alpha$ seems out of reach. Yet, the box potential, where the random walker moves freely between hard walls at $\pm L$, is a useful model system
that presents the whole range of phenomena observed generically.
A key aspect lies in the choice of the jump distribution scaling function
$f(z)$, that can lead to any of the two scenarios mentioned in the  main text: \textcircled{a} a gapped spectrum for which the discrete branch $\lambda_1$ is above the singular continuum, for all jump amplitudes $a$ (regular case, where the singular continuum, although present,
does not play a role in the long time error, and there is no localization); \textcircled{b} a gapless spectrum where the singular continuum becomes the dominant 
relaxation mode for $a>a^*$. This situation \textcircled{b} where localization
appears is the generic case. 
This is why we
focused on case \textcircled{b} in the main text. 

It is useful here to introduce the late-time rejection probabilities $R_a(0)$
at $x=0$, and $R_a(\text{edge})$ at the system's edge, meaning $x=\pm L$
in the box case.
%or $|x|\to \infty$ for a potential in $|x|^\alpha$.
Both depend on $a$. For $a\to 0$, we have $R_a(0)<R_a(\text{edge})$: all moves from $x=0$ are accepted (vanishing rejection probability), while only half of them are, starting from the edge (both in the box case, and when exponent
$\alpha>1$, leading to a convex-up confining potential). A careful inspection of all the numerical data we gathered shows that case \textcircled{a}
corresponds to $R_a(0)<R_a(\text{edge})$ for all $a$;
\textcircled{b} is for the situation where $R_a(0)$ and $R_a(\text{edge})$
do cross for $a=a^*$, so that $R_a(0)> R_a(\text{edge})$ for $a>a^*$. It is then straightforward to realize that 
the behavior of $w(\eta)$ at small $\eta$ discriminates
the two regimes: 
if $w(\eta)$ decreases when increasing $|\eta|$, we have case \textcircled{a}; if $w(\eta)$ increases when increasing $|\eta|$, we have case \textcircled{b}. We considered the family of polynomial $w$-functions, for instance piecewise linear or quadratic
such as 
\begin{align}
w^{(1)}(\eta) &= \frac{1}{a (2 b + c)} \left(b + c \frac{a-|\eta|}{a} \right) \theta(a-|\eta|)\\
w^{(2)}(\eta) &= \frac{1}{a(2 b + 4 c/3)}\left(b + c \frac{a^2-\eta^2}{a^2} \right) \theta(a-|\eta|)
\label{eqw12}
\end{align}
parameterized by the constants $b$ and $c$, in addition to the jump 
size $a$: positive values of $c$ define convex-down functions,
pictorially written $w^\cap(\eta)$ and associated to case 
\textcircled{a}; $c<0$ defines convex-up functions,
denoted $w^\cup(\eta)$, associated to case \textcircled{b}. We take hereafter 
$L=1$, without loss of generality.

\subsection{Numerical results}

\begin{figure}[h]
 \includegraphics[clip=true,width=18cm]{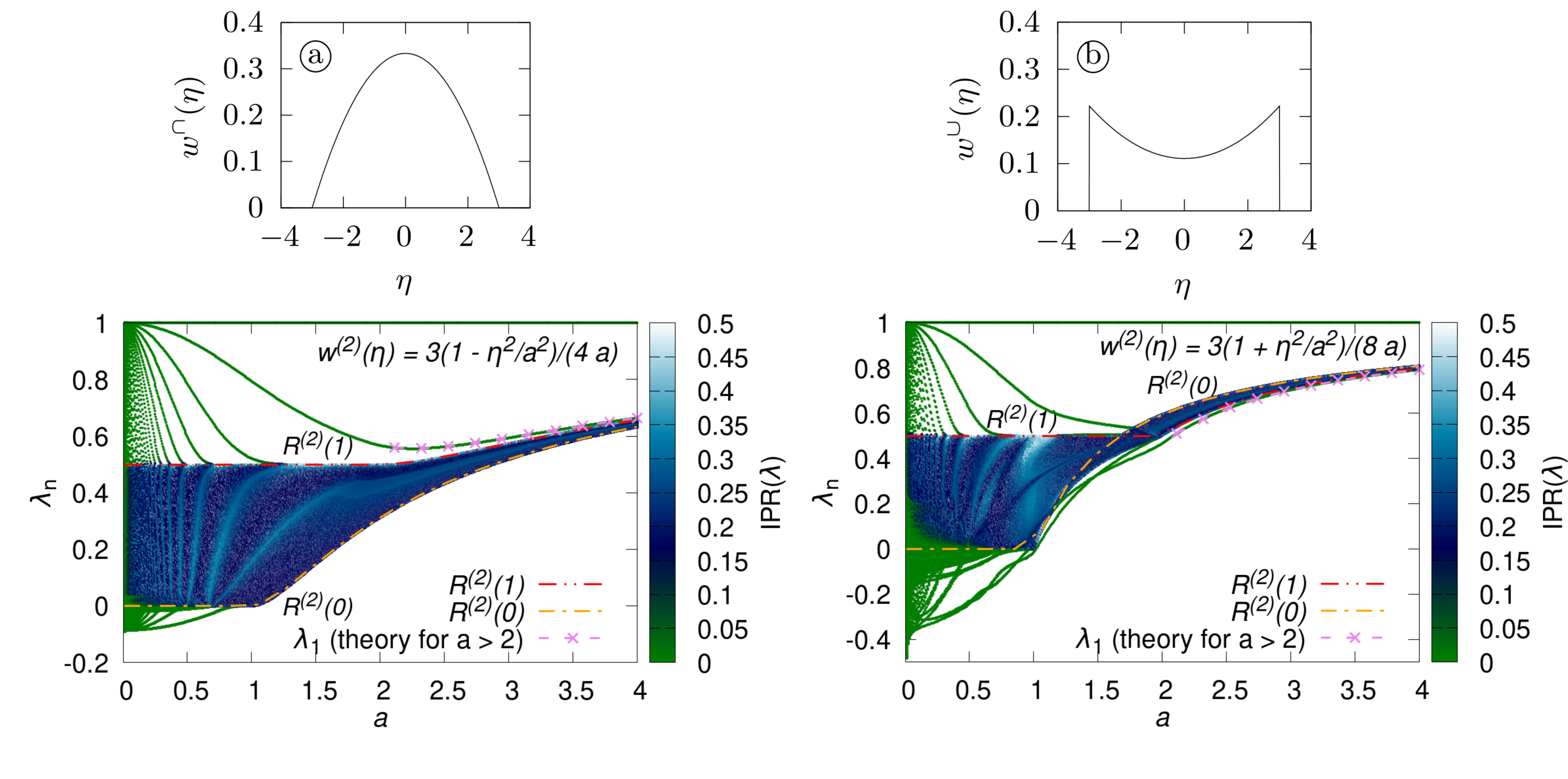} 
\caption{Spectrum of the kernel $F_\beta(x,x')$ for the box potential, with either \textcircled{a} $w^{(2)}(\eta) = w^\cap(\eta)$,
left column ($b=0, c=1$),
or \textcircled{b} $w^{(2)}(\eta) = w^\cup(\eta)$, right column
($b=2, c=-1$). In both cases,
the singular continuum appears in dark blue.
In the gapped case \textcircled{a} where there is no transition, it remains below the discrete $\lambda_1$ branch; the optimal jump can be found by minimizing $\lambda_1$ in Eq.~(\ref{lambda1w2}) which gives $a_{\text{opt}} = \sqrt{c(1+ 3 k^{(2)})(b + c)^{-1}} \simeq 2.251$.
In the gapless case \textcircled{b} on the right,
the singular continuum becomes the dominant relaxation mode for $a>a^*$. Localization ensues,
for $a>a^*=a_{\text{opt}} \simeq 1.79$.
Lengths are in units of the box size ($L=1$).
}
\label{fig:capcup_spectrum}
\end{figure}

%We consider two forms of the jump distribution function for particles confined in $x \in (-1,1)$ with a flat potential $U(x) = 0$ inside the box:

We show in Fig.~\ref{fig:capcup_spectrum} the spectrum of $F_\beta(x,x')$ obtained 
by numerical diagonalization with the parabolic jump distribution $w^{(2)}(\eta)$, either of the type $w^\cap$ or $w^\cup$. The distinction between the
gapped  (\textcircled{a}, with $w^\cap$) and gapless (\textcircled{b}, with $w^\cup$) cases appears. 
At variance with case \textcircled{a}, \textcircled{b}
shows a regime for $a>a^*\simeq 1.79$ where the singular continuum
defines the dominant relaxation mode, so that localization ensues.
The crossing of the curves $R_a(0)$ and $R_a(\text{edge})$ for
$a$ slightly below $a^*$ is also visible. 
For the present box potential,
$U(x)$ either vanishes inside the box, or diverges outside.
Hence, the value of inverse temperature
is irrelevant. We have checked that the qualitative
results remain unchanged for all monotonous (for $\eta > 0$) jump distributions
$w(\eta)$ even in $\eta$, in particular using the piecewise linear distribution 
$w^{(1)}(\eta)$.
Thus for the box potential $U(x) = 0$ ($x \in [-1,1]$) 
 the presence or absence of localization is determined by whether $w(\eta)$ is either 
minimum or maximum at $\eta=0$.

%Furthermore, one may consider that the  present study, for a given jump distribution $w(\eta)$, corresponds to the large temperature limit of the
%spectrum attached to an arbitrary potential $U(x)$, as long as a region 

\subsection{Exact results on eigenvalues of the Monte-Carlo Master equation for a box potential}

We have seen above that the box potential subsumes the gapless/gapped spectra dichotomy, corresponding to the \textcircled{a} absence/\textcircled{b} presence of localization. Besides, the shape of the 
gapless spectrum shown in Fig.~\ref{fig:capcup_spectrum}
is closely reminiscent of its counterpart presented in the main text.
We thus take advantage of the fact that exact results
can be obtained with the box confinement, to shed new light
on the localization phenomenon and its scaling properties.

For a box potential $U(x) = 0$ ($x \in [-1,1]$) the eigenvalue problem of the Monte-Carlo Master equation simplifies into:
\begin{equation}
\lambda \Psi_\lambda(x)= \int_{-1}^1 \Psi_\lambda(y) \,w(x-y)\, dy + \left[1- \int_{-1}^1 w(y-x)\,dy\right]\, \Psi_\lambda(x)
\label{fp.1}
\end{equation}
where $\lambda$ is the eigenvalue and $\Psi_\lambda(x)$ is the eigenvector. For $a > 2$ and the two choices $w^{(p)}(\eta)$ ($p=1,2$) from Eq.~(\ref{eqw12}), this equation reduces to a second order differential equation which can be solved to yield a single eigenvalue $\lambda < 1$. This eigenvalue can be written in the form:
\begin{align}
\lambda^{(p)}_1 = R^{(p)}(1) + (k^{(p)}-1)[R^{(p)}(1) - R^{(p)}(0)].
\label{eqlambdan}
\end{align}
Here, the rejection probability $R^{(p)}(x)$, with index $p = 1$ and $p = 2$ 
%correspond to the two cases in Eq.~(\ref{eqw12}), gives 
is given by
\begin{align}
R^{(p)}(x) = 1 - \int_{-1}^{1} w^{(p)}(x-y) dy .
\label{eqSn}
\end{align}
The constants $k^{(p)}$ 
%take their respective values 
read $k^{(1)} \simeq 1.439$ and $k^{(2)} \simeq 1.356$;
they are the solutions of 
\begin{align}
\frac{1}{\sqrt{k^{(1)}}} \, {\rm arccoth} \sqrt{k^{(1)}} = 1 \;,\; \sqrt{k^{(2)}} 
\, {\rm arccoth} \sqrt{k^{(2)}} = \frac{3}{2} .
\label{eqarccoth}
\end{align}
In both cases, $k^{(p)} > 1$, which implies that if $R^{(p)}(1) > R^{(p)}(0)$, the eigenvalue $\lambda_1^{(p)} > R^{(p)}(1) = \max_x R^{(p)}(x)$ (case \textcircled{a}). On the contrary, if $R^{(p)}(1) < R^{(p)}(0)$, $\lambda_1^{(p)} < R^{(p)}(1) = \min_x R^{(p)}(x)$. 
Thus, it is indeed the comparison between $R^{(p)}(1)$ and $R^{(p)}(0)$ which determines if $\lambda_1^{(p)}$ is above or below the singular continuum, thereby discriminating between \textcircled{a} and \textcircled{b}.

Evaluating the integrals in Eq.~(\ref{eqSn}) we find the explicit expressions (we remind that they are valid for $a>2$):
\begin{align}
\lambda_{1}^{(1)} &= \frac{-2 a b + 2 a^2 b + c - 2 a c + a^2 c + k^{(1)} c}{a^2 (2 b + c)} \label{lambda1w1} \\
\lambda_{1}^{(2)} &= \frac{-3 a^2 b + 3 a^3 b + c - 3 a^2 c + 2 a^3 c + 3 k^{(2)} c}{a^3 (3 b + 2 c)} .
\label{lambda1w2}
\end{align}

%when $w(x)$ are non-degenerate whenever $c \ne 0$. 
To summarize at this point,
for both parametrizations of the jump distribution function and for $a > 2$, there is a single eigenvalue $\lambda < 1$
(besides the singular continuum). This eigenvalue lies above $\max_x R(x)$  or below $\min_x R(x)$ depending on whether $w(\eta)$ is \textcircled{a} maximum or \textcircled{b} minimum at $\eta = 0$. 
Considering that the interval $(\min_x R(x), \max_x R(x))$ is actually filled with singular eigenvalues $\lambda = R(x)$, we describe this situation as $\lambda_1$ lying above or below the singular continuum.

\subsection{Exact results on the localization of the error $\delta P_n(x)$}

We wish to describe analytically the relaxation of the error $\delta P_n(x)$ when $\lambda_1$ lies below $\max_x R(x)$. 
% {\bf not relevant right now but a bit below?}
We remind that for the two parametrizations of $w(\eta)$ from the previous section, a stronger result holds and that in this case, $\lambda_1 < \min_x R(x)$. The Master equation for the error $\delta P_n = P_n - P_\infty$ reads:
\begin{align}
\delta P_{n+1}(x) = \int_{-1}^{1} \delta P_n(y) w(x - y) dy + R(x) \delta P_n(x)
\qquad \hbox{with}\qquad 
R(x) = 1 - \int_{-1}^{1} w(x-y) dy .
\end{align}
Normalization implies $
\int \delta P_{0}(x) dx = \int \delta P_{n}(x) dx = 0$.

% {\bf xxxx assumption of $a$ large enough only relevant from now on}
For $a > 2$ and focusing on the case of a parabolic jump distribution  in Eq.~(\ref{eqw12}), it is possible to simplify notations: 
%writing:
%\begin{align}
%w^{(2)}(\eta) &= w_0 + w_2 \,\eta^2 \\
%R(x) &= r_0 - r_2 \, x^2 \\
%r_0 &= 1 - 2 w_0 - \frac{2 w_2}{3}\;,\; r_2 = 2 \, w_2 ,
%\end{align}
\begin{align}
w^{(2)}(\eta) = w_0 + w_2 \,\eta^2 , \qquad
R(x) = r_0 - r_2 \, x^2 , \qquad
r_0 = 1 - 2 w_0 - \frac{2 w_2}{3} \, , \qquad r_2 = 2 \, w_2 , \quad \hbox{with } w_2>0 .
\end{align}
We then look at symmetric initial conditions $\delta P_0(x) = \delta P_0(-x)$:
\begin{align}
\delta P_{n+1}(x) \, =\,  w_2 \int_{-1}^{1} \delta P_n(y) y^2 dy \, +\,  R(x)\,  \delta P_n(x) . \label{eqrecx2}
\end{align}
We introduce the generating function:
\begin{align}
G(x, z) \, =\,  \sum_{n=0}^{\infty} \delta P_n(x) z^n 
\,= \, \delta P_0(x) + w_2 z \int_{-1}^{1} G(y, z) y^2 dy + z R(x) G(x, z)
\end{align}
from which we get
\begin{align}
  G(x, z) &= \frac{1}{1 - z R(x)} \left( \delta P_0(x) + w_2 z \int_{-1}^{1} G(y, z) y^2 dy \right) .
\end{align}
We then solve for 
$\displaystyle
 G_2(z) = \int_{-1}^{1} G(x, z) x^2 dx .
$
Integrating the Master equation we find:
\begin{align}
G_2(z) &= \frac{\sqrt{r_2 z }}{\sqrt{1 - r_0 z} \arctan \frac{ \sqrt{r_2 z }}{\sqrt{1 - r_0 z}} } \int_{-1}^{1} \frac{\delta P_0(x) x^2}{1 - z R(x)} dx .
\label{eqdefG2}
\end{align}

\begin{figure}[h]
\begin{tabular}{cc}
\includegraphics[clip=true,width=11cm]{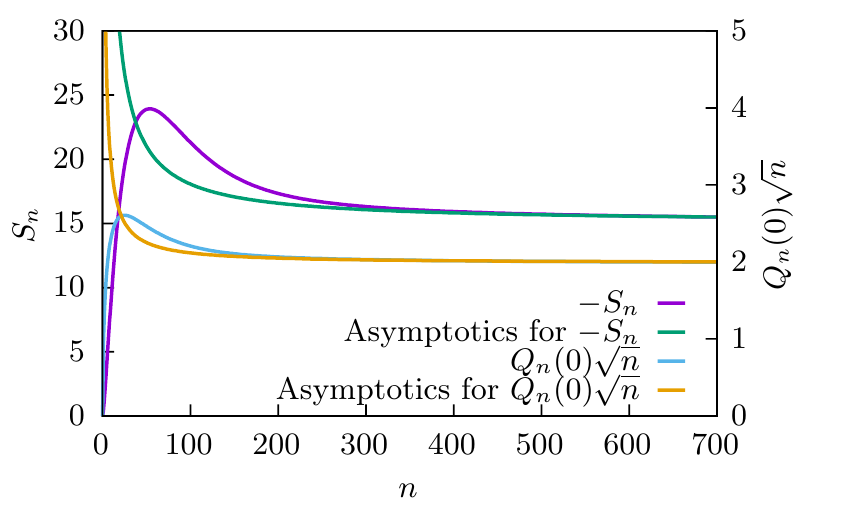} 
\end{tabular}
\caption{Comparison of the numerical results for $S_n = \int_{-1}^{1} \delta P_n(x) x^2 dx$ and $Q_n(x) = R(x)^{-n} \delta P_n(x)$ with the asymptotic estimates in Eqs.~(\ref{eqLimX2},\ref{eqLimP0}). We chose $w^{(2)}(\eta) = 3(1 + \eta^2/a^2)/(8 a)$ and $a = 2.1$; the number of sites for the discretization of the Master equation was 
$N_d = 2\times10^5$. The initial conditions were $\delta P_0(x) = 2 \theta(1/2 - |x|) - 1$ ($r = 1/2$). We note that even if simulations with a finite $N_d$ cannot reproduce the asymptotic power in the $n \rightarrow \infty$ behavior (because of the discrete spectrum), the agreement at finite but large $n$ is nevertheless very good. 
}
\label{FigP0X2VsN}
\end{figure}

To make further progress, we choose as an initial condition
\begin{align}
\delta P_0(x) = r^{-1} \theta(r - |x|) - 1 ,
\label{eqP0sym0}
\end{align}
which allows to compute the integral in 
Eq.~\eqref{eqdefG2} explicitly: 
\begin{align}
G_2(z) = \int_{-1}^{1} G(x, z) x^2 dx \,= \, \frac{2}{r_2 z}\left(1 - \frac{1}{r} \frac{\arctan r Z}{\arctan Z} \right) \qquad 
\hbox{with} \qquad
Z \,= \, \sqrt{ \frac{r_2 z}{1 - r_0 z} } .
\end{align}
We note that $G_2(z)$ is the generating function for the series
\begin{align}
    S_n = \int_{-1}^{1} \delta P_n(x) x^2 dx
\end{align}
which can be viewed as the error in the variance $x^2$ at step $n$. The general method of singularity analysis \cite{acombinatorics} allows to find the asymptotic behavior of a series from the analysis of the singularities of its generating function in the complex plane which are nearest to the origin $z=0$. For $G_2(z)$ the singularity closest to the origin is $z = r_0^{-1}$. The asymptotic expansion of the generating function near this singularity allows us to find:
\begin{align}
% S_n &= \int_{-1}^{1} \delta P_n(x) x^2 dx \\
S_n &\simeq -\frac{2(1-r)}{r^2} \frac{r_0^{3/2}}{(\pi r_2)^{3/2}} \frac{r_0^n}{n^{3/2}} - \frac{(1 - r) r_0^{3/2} [ -48 r^2 r_0 + 4 \pi^2 (1 + r + r^2) r_0 - 15 \pi^2 r^2 r_2 ]}{4 \pi^{7/2} r^4 r_2^{5/2}} \frac{r_0^n}{n^{5/2}} .
\label{eqLimX2}
%  \int_{-1}^{1} P_n(x) x^2 dx &\simeq -\frac{2(1-r)}{r^2} \left( \frac{r_0}{\pi n r_2} \right)^{3/2} r_0^n 
\end{align}
We then introduce the functions $Q_n(x)$ as
\begin{align}
\delta P_n(x) &= R(x)^n Q_n(x) ,
\label{eqdefQn}
\end{align}
and the recurrence equation Eq.~(\ref{eqrecx2}) becomes:
\begin{align}
Q_{n+1}(x) &= \frac{w_2}{R(x)} \int_{-1}^{1} \delta P_n(y) R(x)^{-n} y^2 dy + Q_n(x) \\
&=\frac{w_2}{R(x)} \sum_{m = 0}^n \int_{-1}^{1} P_m(y) R(x)^{-m} y^2 dy + Q_0(x) .
\end{align}
Taking the limit $n \rightarrow \infty$ and 
under the proviso that the series converges,
we find 
\begin{align}
Q_{\infty}(x) &= \frac{w_2}{R(x)}  G_2( R(x)^{-1} ) + \delta P_0(x) 
\end{align}
where $G_2(z)$ is defined in Eq.~\eqref{eqdefG2}. The problem with this expression is that $R(x)^{-1} \ge r_0^{-1}$ lies outside the radius of convergence $|z| \le r_0$ of $G_2(z)$, so this formula is valid only at $x = 0$ when $R(0) = r_0$. Using the obtained value of $G_2(r_0^{-1})$, we find 
\begin{align}
Q_{\infty}(0) = 0 .
\end{align} 
We can then obtain an asymptotic estimate for $Q_n(0)$ :
\begin{align}
Q_{n}(0) &= Q_{\infty}(0) - \frac{r_2}{2 r_0} \sum_{m = n}^\infty \int_{-1}^{1} \delta P_m(y) r_0^{-m} y^2 dy \\
&\simeq \frac{1-r}{r^2 } \frac{r_0^{1/2}}{\pi^{3/2} r_2^{1/2}} \frac{4 n + 1}{2 n^{3/2}} +  \frac{(1 - r) r_0^{1/2} [ -48 r^2 r_0 + 4 \pi^2 (1 + r + r^2) r_0 - 15 \pi^2 r^2 r_2 ]}{12 \pi^{7/2} r^4 r_2^{3/2}} \frac{r_0^n}{n^{3/2}} .
\label{eqLimP0}
\end{align}
From Eq.~(\ref{eqdefQn}), it follows that $\delta P_n(0) = r_0^n Q_n(0)$ where we used $R(0) = r_0$. Equation~(\ref{eqLimP0}) indicates that $Q_n(0)$ decays as a power law 
$n^{-1/2}$ for large $n$:
\begin{align}
\delta P_n(0) \simeq  \frac{2(1 - r)}{\pi^{3/2} r^2} \sqrt{ \frac{r_0}{r_2} } \frac{r_0^{n}}{n^{1/2}} .
\label{eqP0inf}
\end{align}
Comparison of Eqs.~(\ref{eqLimX2},\ref{eqLimP0}) with numerical simulations of discretized approximation of the Master equation are shown in Fig.~\ref{FigP0X2VsN}.

For $x \ne 0$ the series becomes diverging and $Q_\infty(x)$ does not exist. The leading asymptotic behavior can be extracted from the singular behavior of $Q(x, z)$ near $z = r_0^{-1}$:
\begin{align}
\delta P_n(x) \simeq -\frac{1-r}{r^2 x^2} \left( \frac{r_0}{\pi r_2} \right)^{3/2} \frac{r_0^n}{n^{3/2}} .
\label{eqLimPnX}
\end{align}
Interestingly, we find that the ratio $\delta P_n(x)/\delta P_n(0)$ (for $x \ne 0$) does not decay exponentially but as a power law $n^{-1}$.
Comparing Eqs.~(\ref{eqP0inf}) and (\ref{eqLimPnX}), we thus proved the main property of the localizing contribution to the error $\delta P_n(x)$:
\begin{align}
\lim_{n \rightarrow \infty} \delta P_n(x) / \delta P_n(0)  =  0 \quad {\rm whenever} \quad x \ne 0 .
\end{align}

\begin{figure}[h]
\begin{tabular}{cc}
\includegraphics[clip=true,width=9cm]{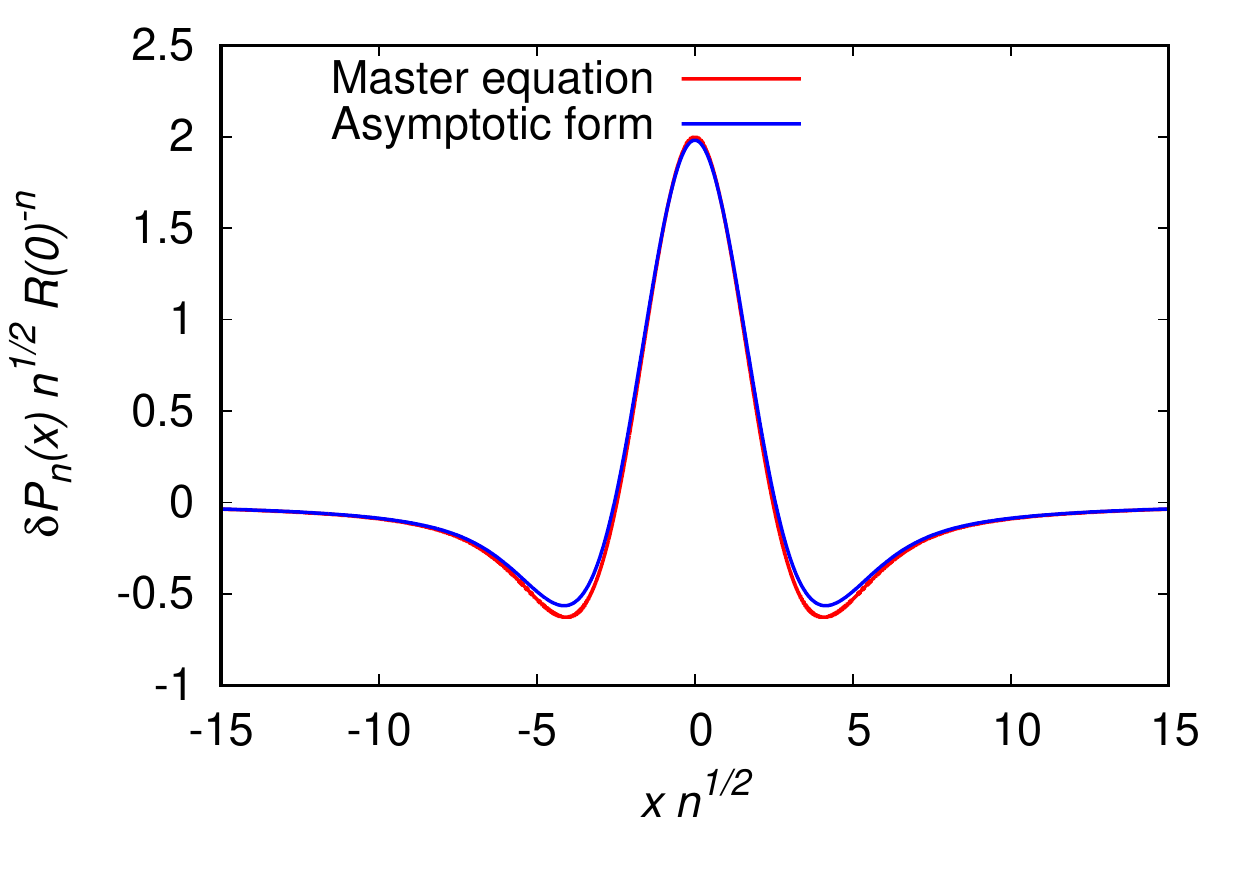} & \includegraphics[clip=true,width=9cm]{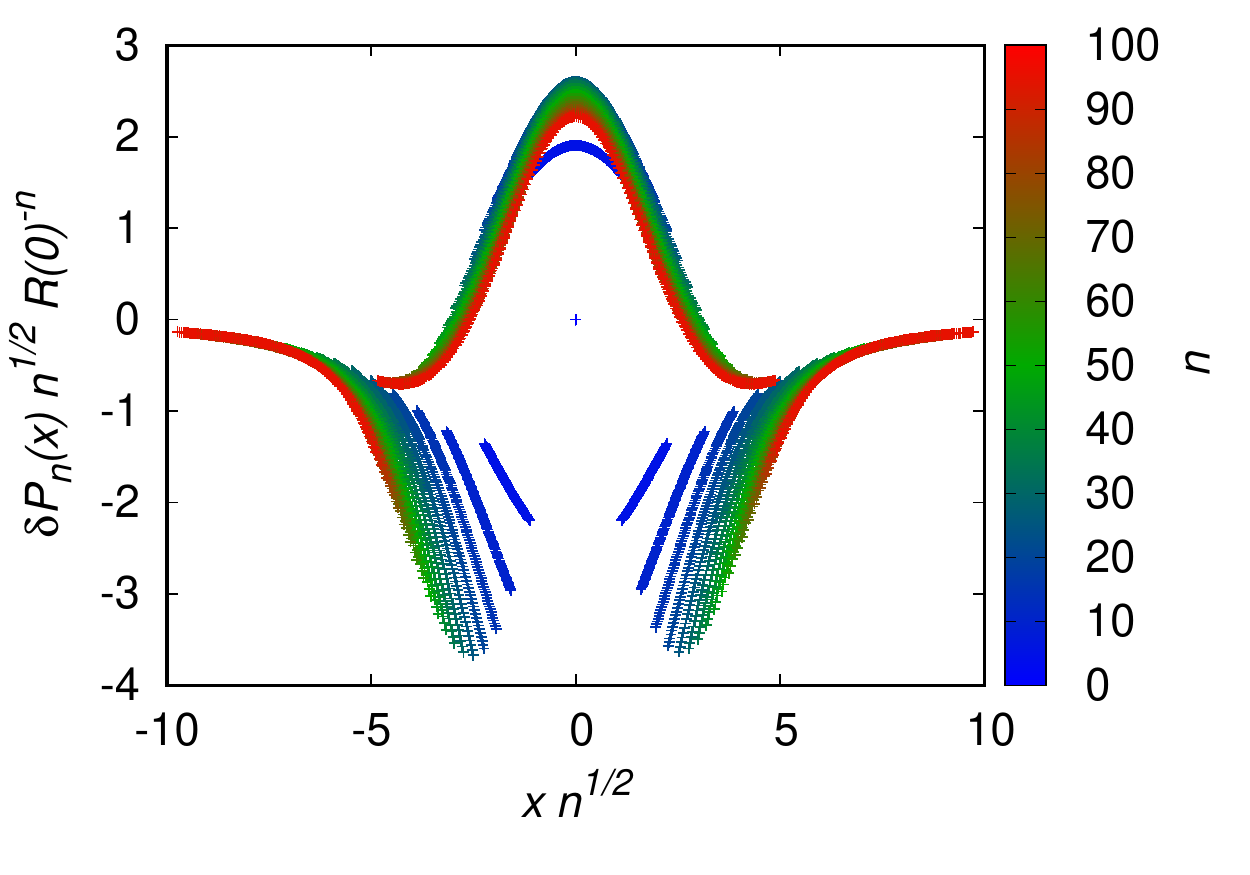} 
\end{tabular}
\caption{Relaxation of $\delta P_0(x) = 2 \theta(1/2 - |x|) - 1$ (i.e. $r = 1/2$) for $w^{(2)}(\eta) = 3(1 + \eta^2/a^2)/(8 a)$ and $a=2.1$; same parameters as Fig.~\ref{FigP0X2VsN}. This figure compares the rescaled $\delta P_n(x)$ as obtained by direct iteration of the Master equation at step $n = 700$, with  the scaling prediction of Eqs.~(\ref{eqScalingX2}) and \eqref{eqScalingPhi}. The right panel shows convergence for $n \le 100$; the moving discontinuity is a trace of the initial distribution $P_0(x)$, that is discontinuous at $x=\pm 1/2$.
Hence, as time proceeds, the central peak extends further, and ultimately reaches the scaling form shown on the left plot. We also illustrated the convergence to the scaling function Eq.~(\ref{eqScalingPhi}) on Fig.~3 from the main text using the same dataset. 
}
\label{FigPnXLimit}
\end{figure}

To find a uniform approximation to $\delta P_n(x)$, we assume the following scaling form %for $\delta P_n(x)$:
\begin{align}
\delta P_n(x) \, = \, \frac{R(0)^{n}}{\sqrt{n}} \, \varphi(x \sqrt{n}) .
\label{eqScalingX2}
\end{align}
Using Eq.~(\ref{eqLimX2}), we can approximate 
\begin{align}
    r_0^{-n} \left( \delta P_{n+1}(x) - R(x) \delta P_n(x) \right) &= \frac{r_2 r_0^{-n} }{2} \int_{-1}^{1} \delta P_n(y) y^2 dy \simeq -\frac{1-r}{r^2} \sqrt{ \frac{r_0^{3}}{\pi^{3} r_2 n^3}} . \label{eqScalingRHside}
\end{align}
On the other hand using Eq.~(\ref{eqScalingX2}), we find:
\begin{align}
      r_0^{-n} \left( \delta P_{n+1}(x) - R(x) \delta P_n(x) \right) &= \frac{r_0}{\sqrt{n+\epsilon}} \varphi(x \sqrt{n+\epsilon})  - (r_0 - r_2 x^2)  \varphi(x \sqrt{n}) \\ 
      &\simeq \frac{r_2 x^2}{n^{1/2}} \varphi(x \sqrt{n}) - \frac{\epsilon r_0 }{2 n^{3/2}}  \varphi(x \sqrt{n}) + \frac{\epsilon r_0 x}{2 n}  \varphi'(x \sqrt{n}) ,
      \label{eqScalingX2Diff}
\end{align}
where we introduced a formal small expansion parameter $\epsilon = 1$ and expanded to first order in $\epsilon$. Introducing ${\widetilde x} = x \sqrt{n}$ and combining Eqs.~(\ref{eqScalingRHside},\ref{eqScalingX2Diff}), we find a first order differential equation on the scaling function $\varphi({\widetilde x})$:
\begin{align}
\frac{r_0 {\widetilde x}}{2}  \varphi'({\widetilde x}) + r_2 {\widetilde x}^2  \varphi({\widetilde x}) - \frac{r_0 }{2}  \varphi({\widetilde x})  = -\frac{1-r}{r^2} \sqrt{ \frac{r_0^{3}}{\pi^{3} r_2}} .
\label{eqDiffDawson}
\end{align}
Equation~(\ref{eqDiffDawson}) admits a single symmetric solution which can be expressed in a compact form introducing the Dawson function: 
\begin{align}
D_+(x) = e^{-x^2} \int_0^x e^{y^2} dy .
\end{align}
We get
\begin{align}
\varphi({\widetilde x}) = \frac{2(1 - r)}{\pi^{3/2} r^2} \sqrt{ \frac{r_0}{r_2} } \left[ 1 - 2 {\widetilde x} \sqrt{\frac{r_2}{r_0}} D_+\left( {\widetilde x} \sqrt{\frac{r_2}{r_0}} \right) \right] .
\label{eqScalingPhi}
\end{align}
From the results
\begin{equation}
    D_+(0) =0 %\quad \hbox{for} \quad x \to 0 
    \qquad \hbox{and} \qquad D_+(x)
    \sim \frac{1}{2 x} + \frac{1}{4 x^3}
    \quad\hbox{for} \quad x\to \infty,
\end{equation}
we recover Eqs. (\ref{eqP0inf}) and (\ref{eqLimPnX}).
Hence, the scaling assumption (\ref{eqScalingX2}) appears fully consistent.
The comparison between $\delta P_n(x)$ obtained by iteration of the Master equation and the prediction of the scaling form is shown in Fig.~\ref{FigPnXLimit}.

%%%%%%%%%%%%%%%%%%%%%%%%%%%%%%%%%%%%%
%%%%%%%%%%%%%%%%%%%%%%%%%%%%%%%%%%%%%%%%%%%%%%
\section{Analytical calculation of the MC relaxation rate for an Harmonic potential}
\label{sec:analytical}

In this section, we show two examples of analytic calculations in the truncated Schr\"odinger eigenbasis, as introduced in sub-sections \ref{sec:schroedinger},\ref{sec:truncated_schroedinger}.

For a harmonic potential $U(x) = x^2/2$, the (dimensionless)  Schr\"odinger equation reduces to the celebrated eigenvalue equation of a quantum harmonic oscillator: 
\begin{align}
\epsilon_n \psi_n(x) = -\psi_n''(x) + \frac{x^2}{4} \psi_n(x) .
\end{align}
The corresponding eigenfunctions can be expressed through Hermite polynomials $H_n$: 
\begin{align}
\psi_n(x) &= \frac{1}{N_n} e^{-x^2/4} H_n( 2^{-1/2} x ) \\
N_n &= \sqrt{ \int dx H_{n}(2^{-1/2} x)^2 e^{-x^2/2} dx } 
\end{align}
where $N_n$ is the normalization. 
To obtain an approximation (and lower bound) for $\Lambda$, we calculate the matrix elements.  
\begin{align}
K_{nm} = \int dy \; \int dy' \psi_n(y) K_{\beta = 1}(y, y') \psi_m(y') 
\end{align}
where the integral kernel is given by Eq.~(\ref{K_symm_T.1}). 
Here, as in the main text, we have expressed positions 
in units of thermal length which amounts to setting $\beta=1$.
This gives the following expression for $K_{nm}$
\begin{align}
K_{nm} &= \int_{-\infty}^{\infty} dy \; w(y) \int_{-\infty}^{\infty} dx \; \psi_m(x - y/2) \; \psi_n(x + y/2) e^{-|x y|/2}  + \int_{-\infty}^{\infty} dx \; \psi_n(x) \psi_m(x) R(x)
\end{align}
where $R(x)$ is the rejection probability. For sufficiently simple expressions of $w(\eta)$ and low values of indices $n$ and $m$, the integrals can be evaluated analytically. 

For a symmetric potential $U(x)$, the truncated matrix splits into a direct sum of odd-even subspaces, as discussed in \ref{sec:truncated_schroedinger}. The sequence of 
$N_s\times N_s$
truncated matrices built from odd eigenfunctions will be noted $K_o^{(N_s)}$.
%where $N_s$ is the number of lowest order odd eigenfunctions in the truncation. 
For example, the matrix $K_o^{(1)}$ reduces to a single scalar $K_{11}$ while $K_o^{(2)}$ is the $2 \times 2$ symmetric matrix with matrix elements corresponding to $K_{nm}$ for $n,m \in \{1,3\}$; higher order approximations are obtained similarly. Likewise with the even sector: the sequence of $N_s\times N_s$
matrices $K_e^{(N_s)}$ is constructed from even wavefunctions. Since $\psi_0(x)$ is an exact eigenvector for any value of the
jump amplitude $a$, the lowest order $K_e^{(1)}$ is given by the scalar $K_{22}$; the next order $K_e^{(2)}$ is given by $K_{nm}$ for $n,m \in \{2,4\}$ and so forth with increasing order $N_s$.
The relaxation rate is then approximated by 
\begin{align}
\Lambda^{(N_s)}_o =  \max {\rm eigenvalues}(K_o^{(N_s)}) \;\;&,\;\;\Lambda^{(N_s)}_e =  \max {\rm eigenvalues}(K_e^{(N_s)}) \\
    \Lambda^{(N_s)} &= \max \left\{\Lambda^{(N_s)}_o , \Lambda^{(N_s)}_e\right\}.
\end{align}

Below, we considered the case of an harmonic potential $U(x)$ for several possible shapes of $w(\eta)$. In all cases, we found that the following approximation is very accurate:
\begin{align}
    \Lambda \simeq \max\left\{ \Lambda^{(N_s)}, \max_x R(x) \right\} 
    \label{eq:approxLambda}
\end{align}
where $R(x)$ is the rejection probability. This expression is operational even for small values $N_s = 2$, and indistinguishable from numerical diagonalization at $N_s = 6$.

\subsection{Harmonic potential with a flat jump distribution $w(\eta)$}

\begin{figure}[h]
\begin{tabular}{cc}
\includegraphics[clip=true,width=9.5cm]{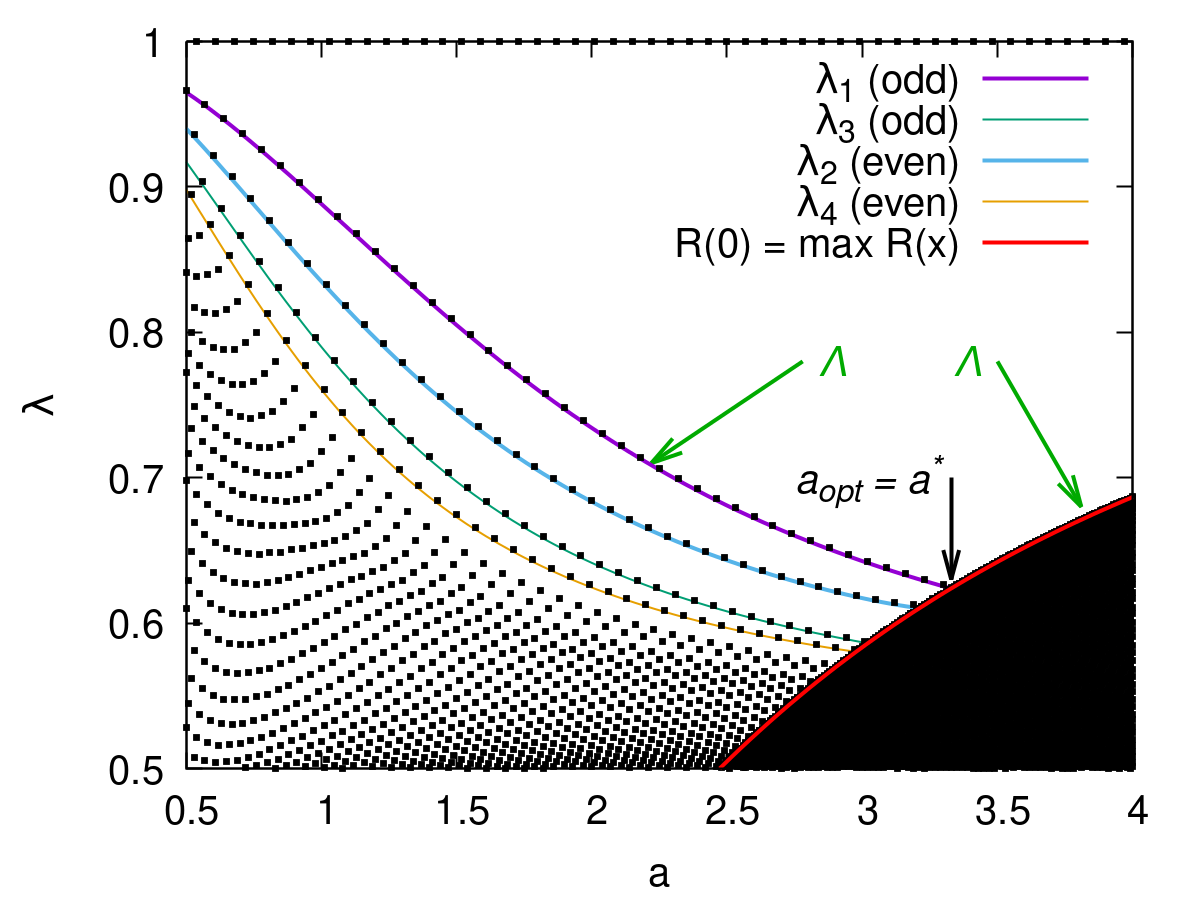} 
\end{tabular}
\caption{Comparison between numerical eigenvalues and  the analytical approximation  for $U(x) = x^2/2$ with flat jump distribution $w(\eta) = \theta(a - |\eta|)/(2 a)$. The symbols show numerical eigenvalues of the discretized Master equation ($N_d = 2000$, $X_{max} = 10$) while the continuous curves display the analytical results for the four slowest relaxation eigenmodes: $\lambda_{1},\lambda_{2},\lambda_{3}$ and $\lambda_{4}$. The analytical approximation is obtained from the two highest eigenvalues of $K_o^{(6)}$ (yielding $\lambda_1$ and $\lambda_3$) and $K_e^{(6)}$ (yielding $\lambda_2$ and $\lambda_4$). The analytical top eigenvalue $\Lambda$ is thus here $\Lambda^{(6)}$.
We see that below $\Lambda$, the other eigenvalues are also indistinguishable from the numerical eigenvalues, until the crossing with the singular continuum, bounded by $R(0) = \max_x R(x)$. This shows that lower eigenvalues are also very accurately reproduced in this approximation, for values of $a$ below the localisation transition of the corresponding mode. The maximum rejection probability is given by Eq. \eqref{eq:RmaxFlat}.}
\label{evalx2step}
\end{figure}

We report explicit results for the lowest order terms for $w(\eta) = \theta(a - |\eta|)/(2 a)$, indicating that the scaling function $f$ reads 
$f(z) = \theta(1 - |z|)/2$. We find: 
\begin{align}
K_{11}(a) &= 1-\frac{a^3 \text{erfc}\left(\frac{a}{2 \sqrt{2}}\right)-2 \sqrt{\frac{2}{\pi }}
   \left(a^2+8\right) e^{-\frac{a^2}{8}}+16
   \sqrt{\frac{2}{\pi }}}{6 a} \\
K_{13}(a) &= \frac{-\sqrt{2 \pi } a^5
   \text{erfc}\left(\frac{a}{2
   \sqrt{2}}\right)+4 \left(a^4+a^2+8\right)
   e^{-\frac{a^2}{8}}-32}{20 \sqrt{3 \pi } a} \\
   K_{33}(a) &= 1-\frac{1}{420} \left(5 a^4+63 a^2+210\right)
   a^2 \text{erfc}\left(\frac{a}{2
   \sqrt{2}}\right)+\frac{\left(20 a^6+207
   a^4+372 a^2+2976\right)
   e^{-\frac{a^2}{8}}}{420 \sqrt{2 \pi }
   a}-\frac{124 \sqrt{\frac{2}{\pi }}}{35
   a}
\end{align}

The steady state rejection probability $R_\infty$ is given by:
\begin{align}
    R_\infty = \frac{2}{a} \sqrt{\frac{2}{\pi }}
   \left(e^{-\frac{a^2}{8}}-1\right)+\text{erf}\left(\frac{a}{2 \sqrt{2}}\right) ,
\end{align}
and the maximum rejection probability reads:
\begin{align}
    R(0) = \max_x R(x) = 1 - \frac{\sqrt{\pi/2}}{a} \text{erf}\left(\frac{a}{\sqrt{2}}\right) .
    \label{eq:RmaxFlat}
\end{align}
This gives explicit expressions for the first two orders:
\begin{align}
\Lambda^{(1)} &= \Lambda_o^{(1)} = K_{11}(a) \\
\Lambda^{(2)} &= \Lambda_o^{(2)} = \frac{K_{11}(a)+K_{33}(a)}{2} + \sqrt{ K_{13}(a)^2 + \left(  \frac{K_{11}(a)-K_{33}(a)}{2}  \right)^2} .
\end{align}
We do not report explicit expressions for higher $K_{nm}$ matrix elements, as expressions become more  cumbersome. From Fig.~4 (in the main text), we see that this approximation quickly converges for $a < a^*$ and that $\Lambda^{(2)}$ is already very close to the value of $\Lambda$ obtained by numerical diagonalization. For $a > a^*$, the convergence of this expansion is much slower and $\Lambda$ is instead given by the maximum rejection probability,
as explained in the main text.

The rapid convergence of the approximation Eq.~(\ref{eq:approxLambda}) with increasing $N_s$ was already illustrated on Fig.~4 from the main text for the slowest relaxation mode $\Lambda$. Here we show that this approximation allows also to obtain accurate expressions for other sub-leading relaxation modes (see Fig.~\ref{evalx2step}).

% cp ~/Cprograms/phys/cooling/simMC/evalx2gauss.png . 
% cp ~/Cprograms/phys/cooling/simMC/evalx2step.png . 
% cp ~/Cprograms/phys/cooling/simMC/evalx2stepx2.png . 

\subsection{Harmonic potential with a Gaussian jump distribution $w(\eta)$}

\begin{figure}[ht]
\begin{tabular}{cc}
\includegraphics[clip=true,width=9cm]{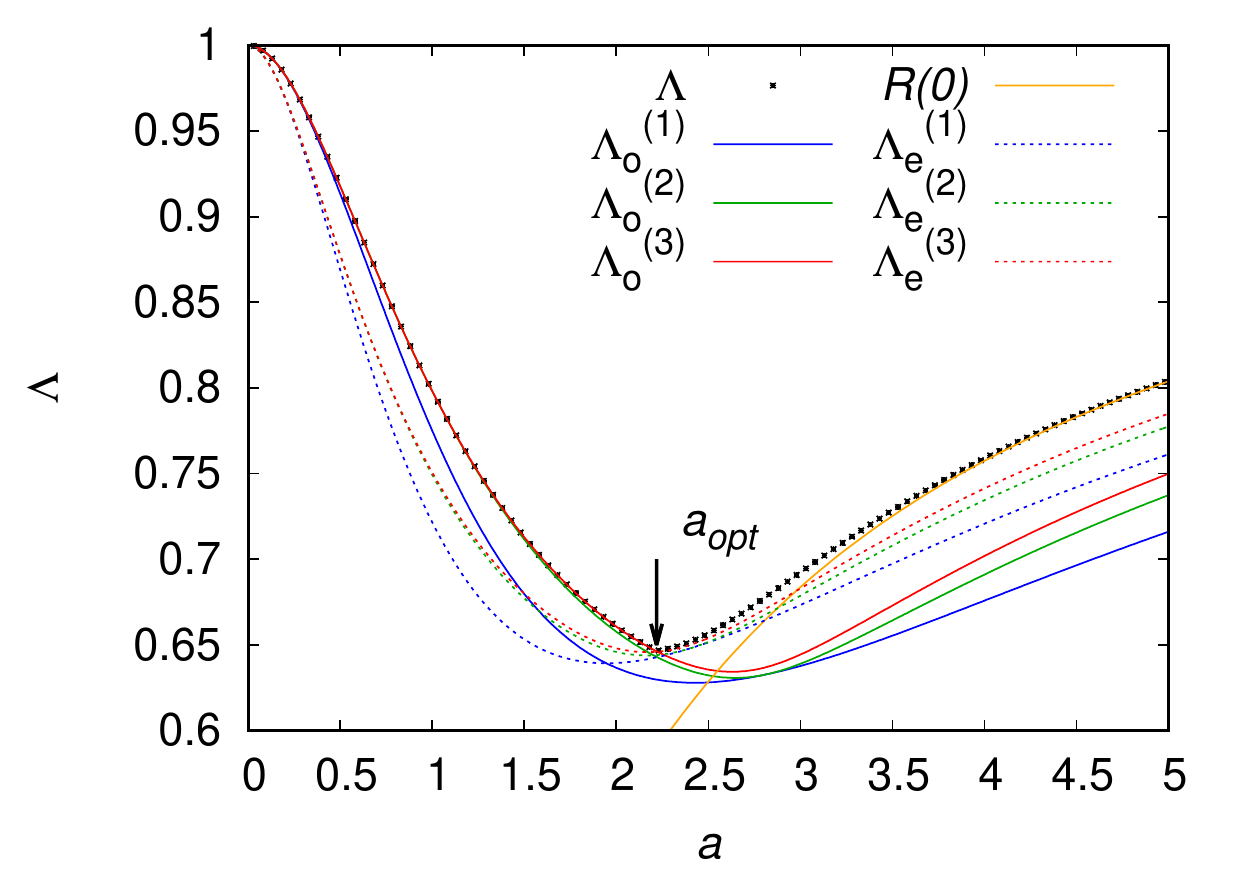} & \includegraphics[clip=true,width=8.3cm]{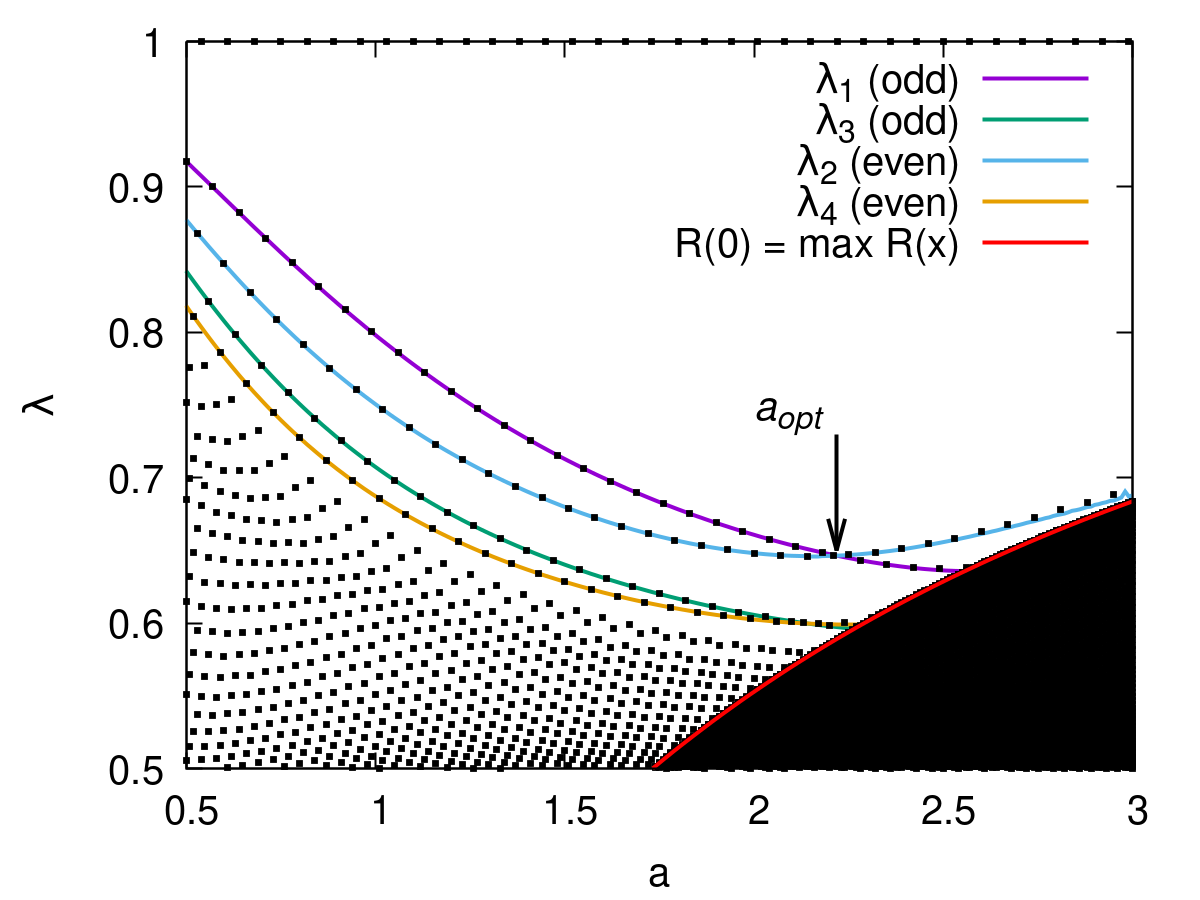}
\end{tabular}
\caption{The left panel shows analytic approximations for  $\Lambda$ for $U(x) = x^2/2$ with a Gaussian jump distribution $f(z) = (\sqrt{2 \pi})^{-1} \exp\left( -z^2/2  \right)$. In this case, the matrices $K^{(N_s)}_o$ provide a good approximation for $\Lambda$ for $a < a_{opt} \simeq 2.21845$. At $a = a_{opt}$, the symmetry of the leading relaxation mode changes from odd to even and for $a > a_{opt}$, $\Lambda$ is instead well approximated by $\Lambda_e^{(N_s)}$, 
becoming increasingly close to the maximum of the rejection probability as $a$ increases further. We note, however, that there does not  seem to be a localization transition in this model: the IPR of the eigenvector for $\Lambda$ follows a $N_d^{-1}$ scaling suggesting that for this case ${\cal N} = 1$. 
The right panel is similar to Fig.~(\ref{evalx2step}), 
%but for the case of Gaussian $w(x)$ 
confirming that $K_o^{(6)}$ and $K_e^{(6)}$ provide very good approximations for the leading eigenvalues up to the crossing with the maximum of the rejection probability, given by Eq. \eqref{eq:RmaxGaussian}. Using this method, we find position of the localization transition for slowest antisymmetric relaxation mode $\lambda_2$ at $a \simeq 2.55657$.
}
\label{evalx2gauss}
\end{figure}

For a Gaussian jump distribution 
\begin{align}
w(\eta) = \frac{1}{a \sqrt{2 \pi}} \exp\left( -\frac{\eta^2}{2 a^2} \right) \quad
\Leftrightarrow \quad
f(z) = \frac{1}{\sqrt{2 \pi}} \exp\left( -\frac{z^2}{2} \right) ,
\end{align}
we find the matrix elements for the odd subspace of the Schr\"odinger eigenbasis:
\begin{align}
K_{11}(a) &= 1-\frac{a^2}{2}+\frac{a^2 }{\pi
   }\arctan\left(\frac{a}{2}\right)+\frac{2 a^3}{\pi  \left(a^2+4\right)} \\
K_{13}(a) &= \sqrt{\frac{3}{2}}\frac{ a^4}{\pi
   } \arctan
   \left(\frac{a}{2}\right)+\frac{\left(12 a^4+80 a^2-3 \pi 
   \left(a^2+4\right)^2 a+96\right) a^3}{2
   \sqrt{6} \pi  \left(a^2+4\right)^2} \\
   K_{33}(a) &= 1 -\frac{\left(5 a^4+9 a^2+6\right) a^2 }{2 \pi
   } \arctan\left(\frac{2}{a}\right)+\frac{\left(15 a^8+187 a^6+834 a^4+1560
   a^2+1152\right) a^3}{3 \pi 
   \left(a^2+4\right)^3}
\end{align}

For the Gaussian jumps, $\Lambda$ also depends on the matrix elements in the even subspace:
\begin{align}
K_{22}(a) &= \frac{2 \left(3 a^3+\pi \right)-a^2 \left(3
   a^2+4\right) \arctan\left(\frac{2}{a}\right)}{2 \pi } \\
   K_{24}(a) &= \frac{a^3}{4
   \sqrt{3} \pi  \left(a^2+4\right)} \left[30 a^4+128 a^2-3
   \left(a^2+4\right) \left(5 a^2+8\right) a
   \arctan\left(\frac{2}{a}\right)+64\right] \\
   K_{44}(a) &= 1-\frac{\left(35 a^6+80 a^4+72 a^2+32\right)
   a^2 \arctan\left(\frac{2}{a}\right)}{8
   \pi }+\frac{\left(105 a^8+940 a^6+2712
   a^4+3072 a^2+1920\right) a^3}{12 \pi 
   \left(a^2+4\right)^2}
\end{align}
We also get:
\begin{align}
R(0) &= \max_x R(x) = 1 - \frac{1}{\sqrt{1+ a^2}}
\label{eq:RmaxGaussian}\\
R_\infty &= 1 - \frac{2}{\pi} \arctan \frac{2}{a}
\end{align}

%\begin{figure}[h]
%\begin{tabular}{cc}
%\includegraphics[clip=true,width=9.5cm]{evalx2stepx2.png} 
%\end{tabular}
%\caption{$U(x) = x^2/2$ with jumps $w(x) = 3 x^2 \theta(a- |x|)/(2 a^3)$}
%\label{evalx2gauss}
%\end{figure}

Figure \ref{evalx2gauss} compares the result of the Schr\"odinger eigenbasis approximation for a Gaussian $w(\eta)$ to numerical eigenvalues for the discretized Master equation. We do not find evidence of a localization transition for $\Lambda$, but instead a change of parity at $a = a_{opt} \simeq 2.21845$. As for the case of a flat jump distribution shown on Fig.~\ref{evalx2step}, the Schr\"odinger eigenbasis approximation works very accurately for all the slowest relaxation modes until they cross the singular continuum. It seems that even if ${\cal N} = 1$ for this case, the maximum rejection probability $\max_x R(x)$ is still a very good approximation for $\Lambda$ at large $a$ ($a \ge 4$).

\begin{figure}[hbt]
\begin{tabular}{cc}
\includegraphics[clip=true,width=9.5cm]{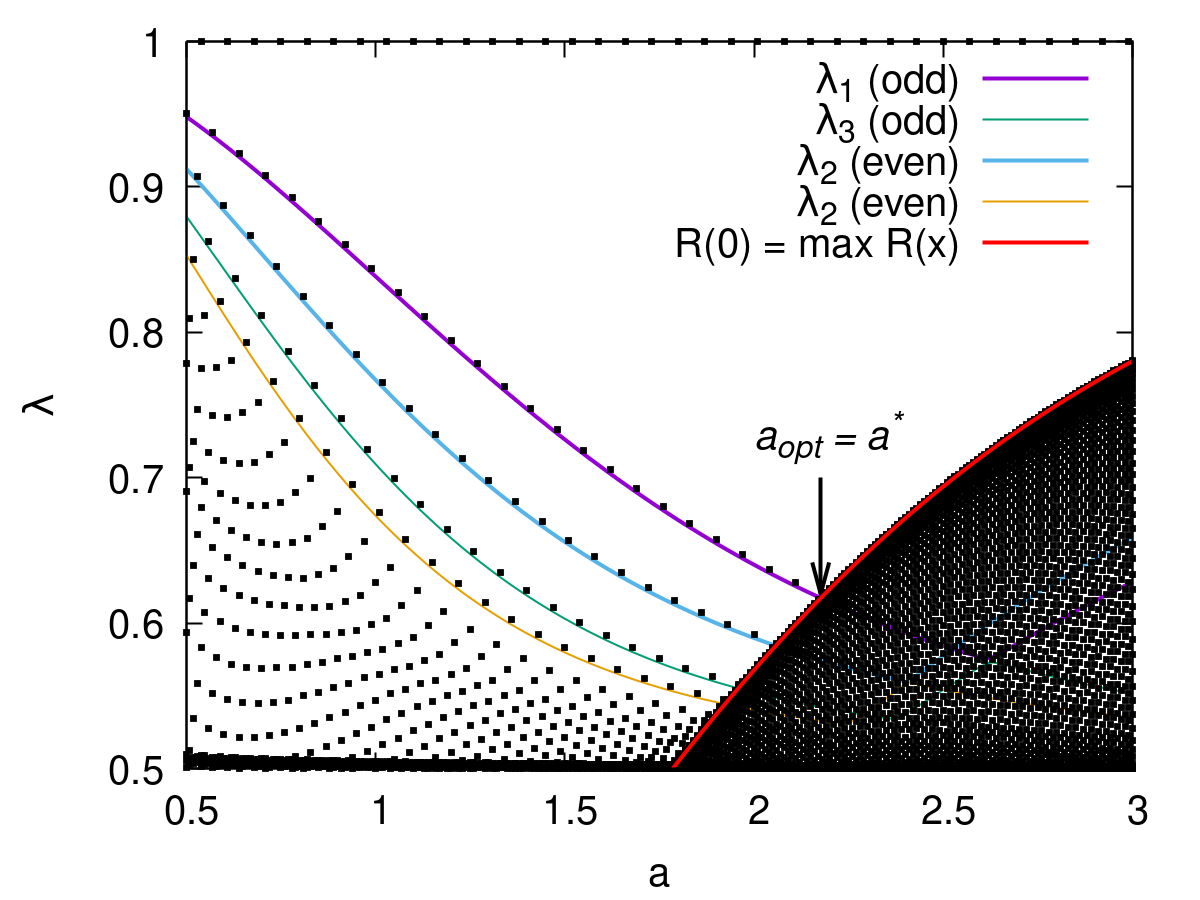} 
\end{tabular}
\caption{Results for $U(x) = x^2/2$ with a jump distribution $w(\eta) = |\eta| \theta(a- |\eta|)/a^2$. The continuous curves show the results of the analytical calculation $K_o^{(6)}$ and $K_e^{(6)}$; they provide a very good approximation for the leading eigenvalues up to the crossing with the maximum of the rejection probability.
}
\label{evalx1gauss}
\end{figure}

\subsection{Harmonic potential with jump distribution $w(\eta) = a^{-2} |\eta| \theta(a-|\eta|)$}

Again for a harmonic potential, analytical results for this shape of $w(\eta)$ can be obtained in the same way as above. We do not report them here, and only provide a comparison between numerical and analytical calculations on Fig.~\ref{evalx1gauss}.

\subsection{Comparing the different jump distributions}

\begin{table}[hb]
\begin{center}
\begin{tabular}{|c|c|c|c|c|}
\hline $w(\eta)$ & $a_{\text{opt}}$ & $a^*$ & \;\; $\Lambda(a_{\text{opt}})$ \;\; & $1-R_\infty(a_{\text{opt}})$ \\ \hline
$(2 a)^{-1} \theta(a-|\eta|)$ & 3.32878 & $a^* = a_{\text{opt}}$ & 0.62382 & 0.45543 \\ \hline
$(a \sqrt{2 \pi})^{-1} \exp(-\eta^2/(2 a^2))$ & 2.21845 & \hbox{none} & 0.64638 & 0.467 \\ \hline
$a^{-2} |\eta| \theta(a-|\eta|)$  & 2.17613 & $a^* = a_{\text{opt}}$ & 0.6172 & 0.482 \\ \hline
\end{tabular}
\end{center}
\caption{Summary of the results on the optimal jump length $a$ for several shapes of the jump distribution $w(\eta)$ in a harmonic potential. These three cases correspond to Figures
\ref{evalx2step}, \ref{evalx2gauss}, \ref{evalx1gauss}}
\label{tableI}
\end{table}

Among the three jump distributions worked out above, the last one provides the value $\Lambda(a_{\text{opt}}) = 0.61723$,
which is the lowest among the studied examples. 
In this respect, this  jump distribution,
at the optimal jump amplitude $a_{\text{opt}}$, yields 
the fastest method for sampling the equilibrium 
distribution. Results are summarized in Table 
\ref{tableI}.

\begin{figure}[h]
\begin{tabular}{cc}
\includegraphics[clip=true,width=9.5cm]{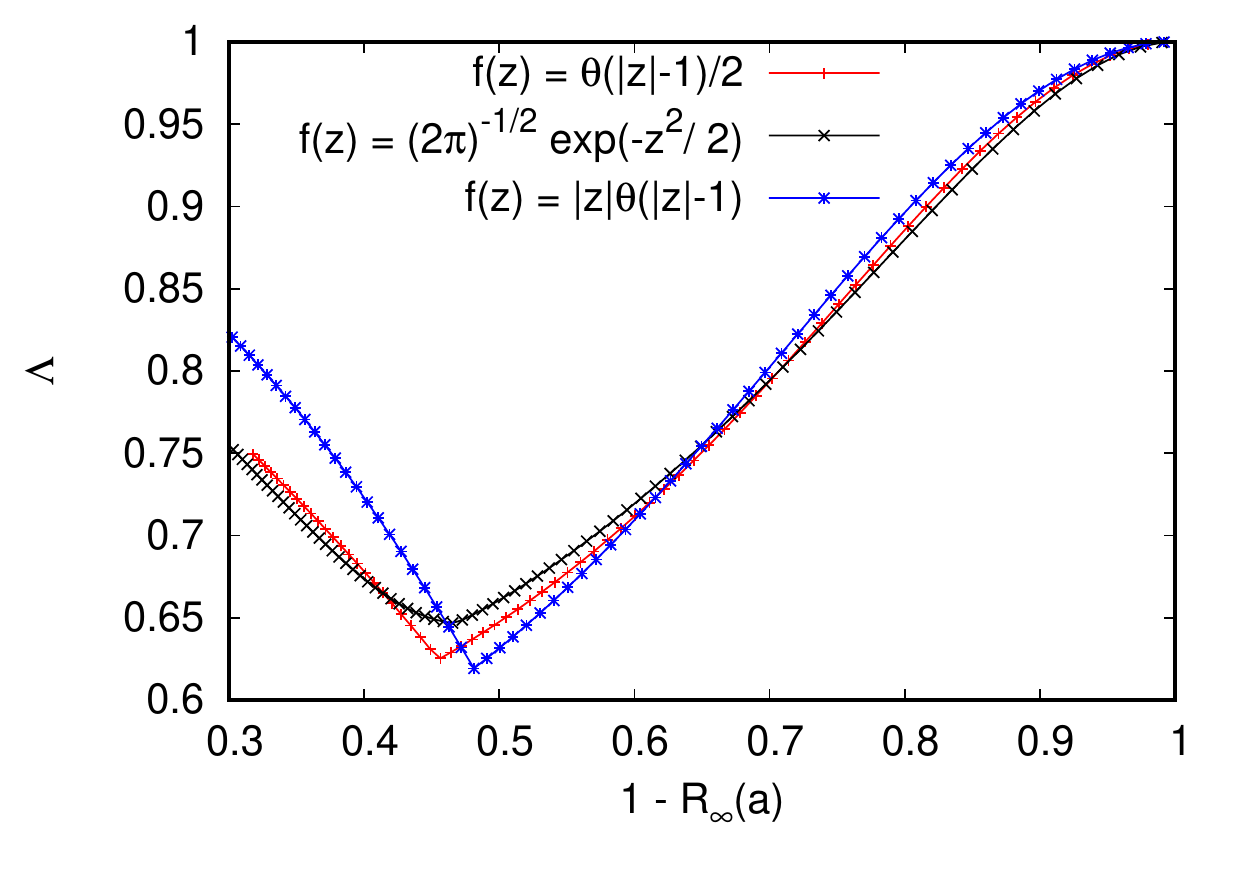} 
\end{tabular}
\caption{Comparing different jump distributions. Plots of the leading relaxation eigenvalue $\Lambda$ as a function of the acceptance probability, for the three cases summarized in Table \ref{tableI}, corresponding to Figures
\ref{evalx2step}, \ref{evalx2gauss}, \ref{evalx1gauss}.
}
\label{LambdaVsAcceptance}
\end{figure}

Figure \ref{LambdaVsAcceptance} compares the spectral results for the three jump distributions. We note that they correspond to a $w(\eta)$ that is either
increasing, flat, or decreasing with $|\eta|$. In spite 
of these differences, the leading relaxation eigenvalue $\Lambda$ displays the same behaviour as a function 
of the acceptance probability $1-R_\infty$. In particular,
the three cases feature optimality (smallest $\Lambda$,
fastest convergence) for an acceptance probability close to 
50\%.

%%%%%%%%%%%%%%%%%%%%%%%%%%%%%%%%%%%%%%%%

\section{Generalization: beyond one dimension and inclusion of interactions}
\label{sec:generalizations}

While the results presented so far focused on one-dimensional dynamics, 
we here put to the test  the generality of the localization transition by
considering more generic models, beyond 1D or 
with interacting degrees of freedom. The analysis is
here mostly numerical.

\subsection{Beyond 1D}

\begin{figure}[h]
\includegraphics[clip=true,width=9cm]{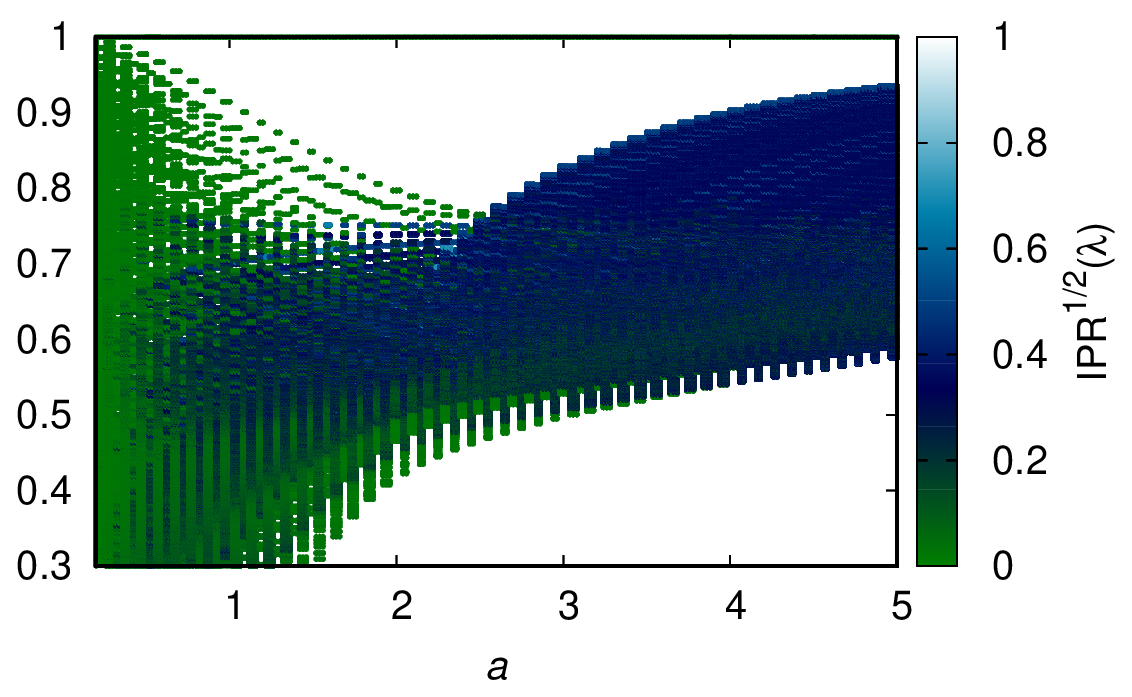} 
\caption{Spectrum of the Monte-Carlo Master equation kernel for a two dimensional particle confined to the cell $(-5,5)\times(-5,5)$ (discretized to $100^2$ boxes) and in the potential $U(x,y) = (x^2+y^2)/2$ ($\beta = 1$). The attempted jumps are two dimensional changing both $x$ and $y$ in an interval $(-a,a)$ centered around their initial values. Color shows IPR$^{1/2}$, where the square root is used to enhance contrast (the lower contrast in IPR values is related to the high symmetry of the potential $U(x,y)$, see for example the higher contrast in Fig.~\ref{FigCC2D} where all symmetries are broken).
}
\label{Fig2D}
\end{figure}

Simulations in higher dimensions rapidly become demanding in terms of numerical resources. In two dimensions, it is still possible to use direct diagonalization to obtain the full eigenspectrum of the Master equation and the IPR of the eigenvectors. An example of such a simulation is shown on Fig.~\ref{Fig2D}: the results are very similar to the one dimensional simulation in Fig.~1 (main text) except that $a^* \simeq 2.6$ instead of $a^* \simeq 3.3$ due to the two dimensional nature of attempted jumps.

\begin{figure}[h]
\includegraphics[clip=true,width=9cm]{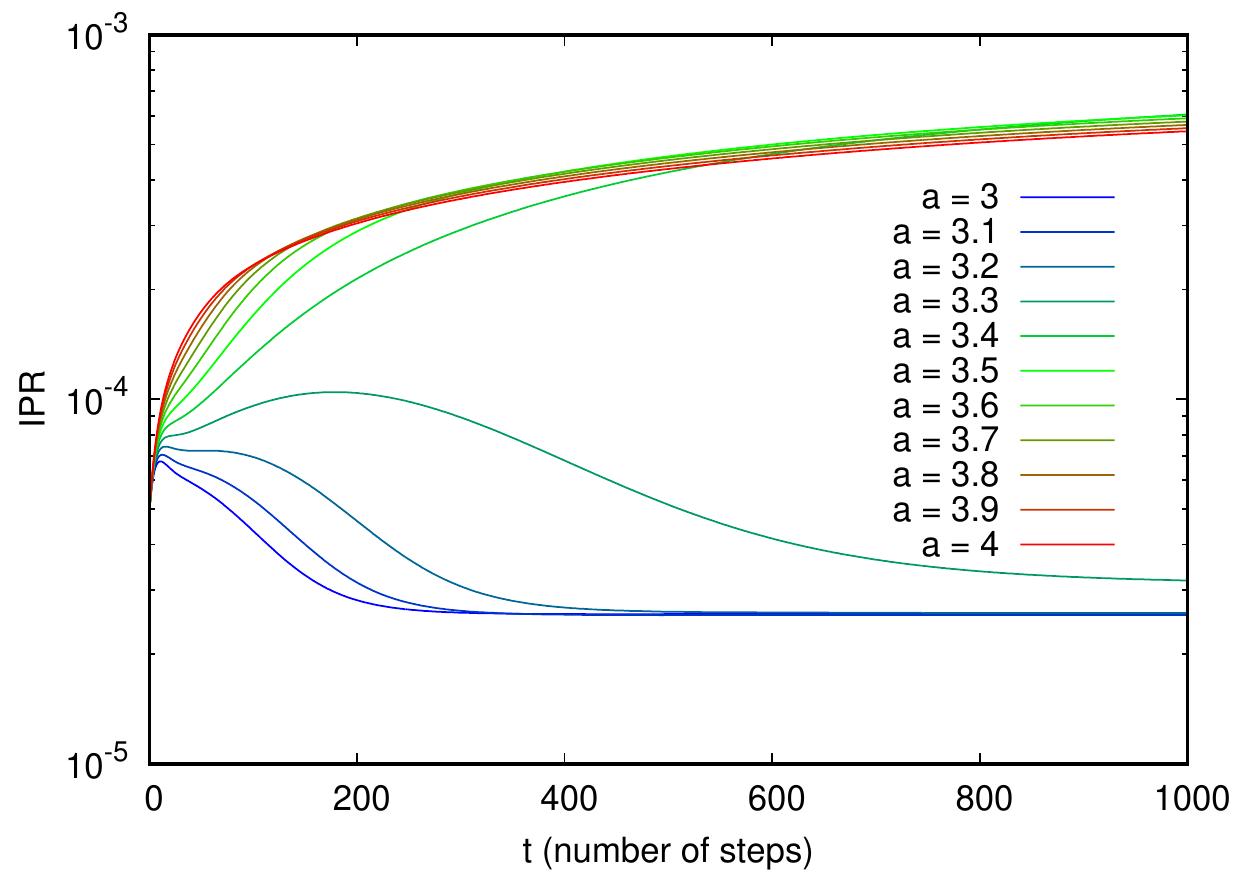} 
\caption{A 3D example with the potential $U(x,y,z) = (x^2+y^2+z^2)/2$ and a confinement volume $(-5,5)^3$ discretized in $100^3$ boxes. The initial distribution $P_0(x,y,z)$ is an off-centred Gaussian.  
}
\label{Fig3D}
\end{figure}

Simulations in 3D are numerically more accessible if jumps are attempted in only one of the directions $x,y,z$ at a time. This makes the matrix representing the Master equation kernel sparse, allowing to find the time evolution of the error distribution $\delta P_n = P_n - P_\infty$. We show
in Fig.~\ref{Fig3D} the evolution of the IPR of $\delta P_n$ with the number of algorithm steps (time). A sharp transition from decreasing to increasing IPR as a function of time is seen around $a = 3.3$. Since the attempted jumps are 1D, the localization transition takes place at the same value as for the 1D harmonic potential.

\subsection{Interactions}

\begin{figure}[h]
\begin{tabular}{cc}
Repulsive $U_+$ & Attractive $U_-$ \\
\includegraphics[clip=true,width=9cm]{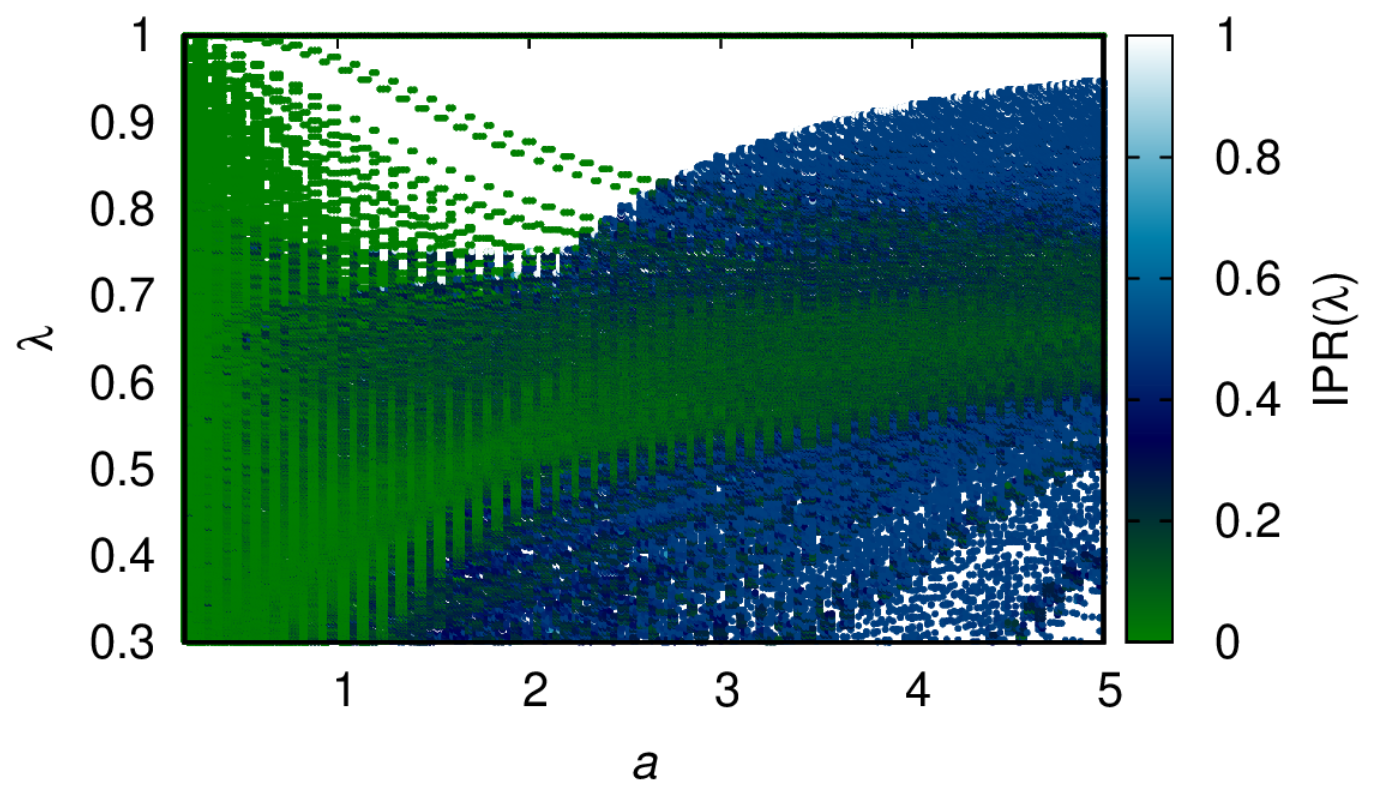} & \includegraphics[clip=true,width=9cm]{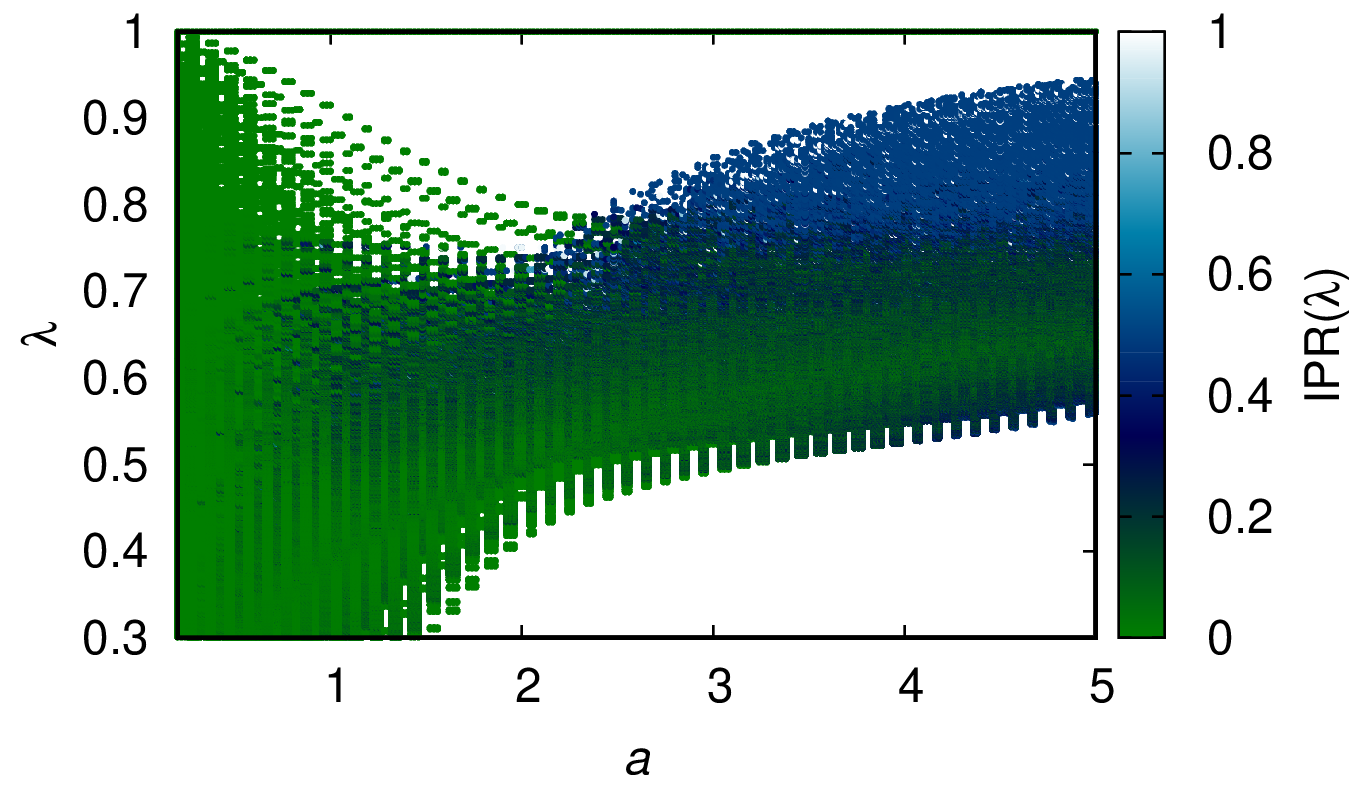} 
\end{tabular}
\caption{Spectrum of the Monte-Carlo Master equation kernel for two interacting particles, with interaction potential given by Eq.~(\ref{eq:MCC2D}). Two situations were investigated, 
with a repulsive (left panel) or an attractive potential (right panel). The configuration space, restricted to the interval $(-5,5)$,  was discretized in $100 \times 100$ cells.
}
\label{FigCC2D}
\end{figure}

We provide a numerical example illustrating  the localization transition in the Monte Carlo relaxation of interacting particles. We consider a case which is numerically tractable by full diagonalization, in analogy with Fig.~1 from the main text and with Fig. \ref{fig:capcup_spectrum}. We consider two particles at positions $x_1$ and $x_2$ in a one dimensional box, with $x_1, x_2 \in [-5,5]$. The energy of a configuration $(x_1,x_2)$ is given by the potential:
\begin{align}
U_\pm(x_1, x_2) = \frac{x_1^2 + x_2^2}{2} \pm \frac{2}{0.1 + |x_1 - x_2|} + x_1 - x_2 
\label{eq:MCC2D}
\end{align}
where, depending on the plus or minus signs, the  interaction between $x_1$ and $x_2$ is repulsive ($U_+$) or attractive ($U_-$). We simulate the steady state of this system using a Monte-Carlo algorithm, with jumps where we attempt to simultaneously change $x_1$ and $x_2$ in an interval $(-a,a)$ around their initial position. The spectrum of the corresponding Master equation is shown in Fig.~\ref{FigCC2D}, indicating that a localization transition occurs in this case even when interactions are present. Switching from repulsive to attractive interaction changes the value of the optimal jump length $a^*$, and the spread of the eigenspectrum. In both cases however, the IPR drastically increases for $a > a^*$, indicating a localization transition.

\subsection{Relaxation in presence of multiple local minima}

\begin{figure}[h]
\begin{tabular}{cc}
\includegraphics[clip=true,width=8.5cm]{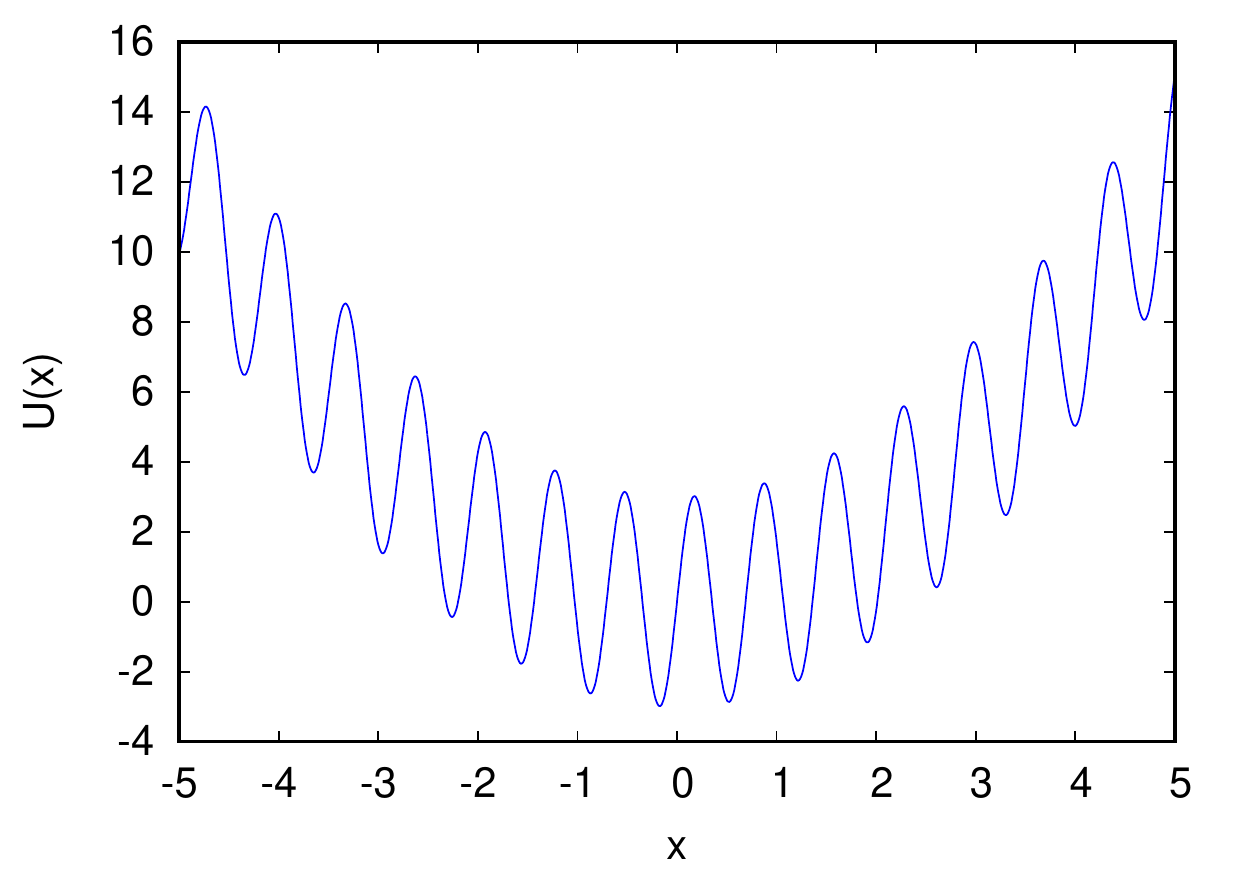} & \includegraphics[clip=true,width=9.5cm]{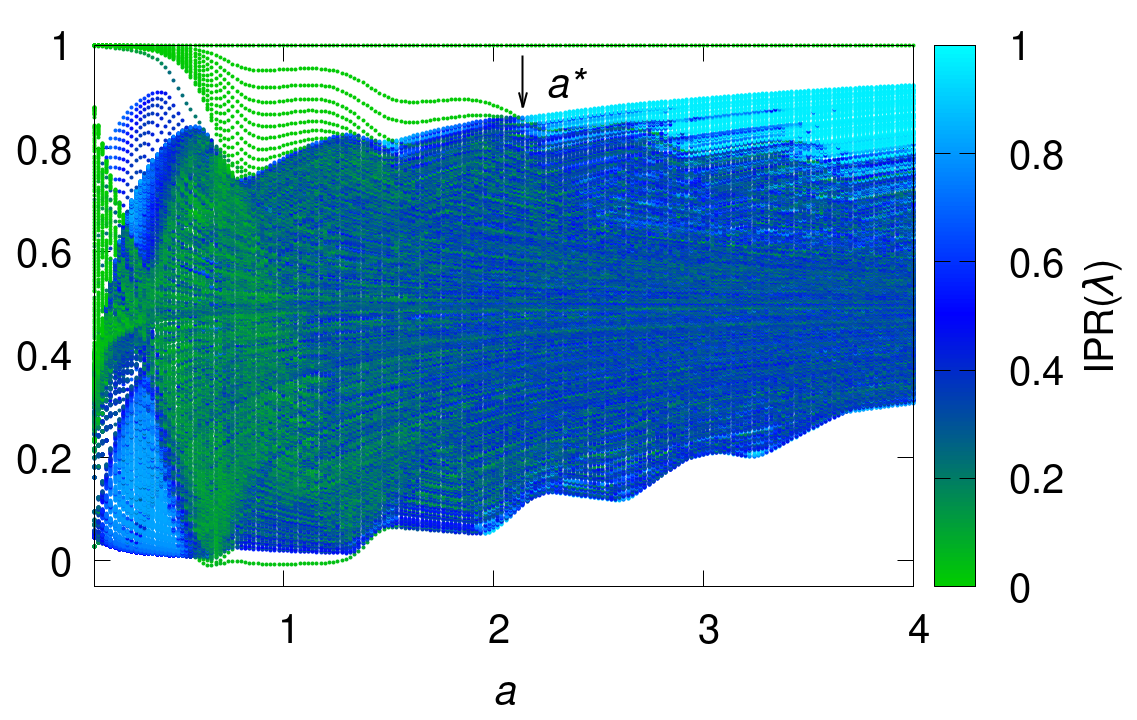} 
\end{tabular}
\caption{(left panel) Example of a potential with many local minima given by Eq.~(\ref{eqmanymin}).
(right panel) Monte Carlo relaxation eigenspectrum for this potential with a flat jump distribution as in Fig.~1 from the main text. The box $(-5,5)$ was discretized in $10^3$ sites.}
\label{FigNmin}
\end{figure}

Finally, we illustrate numerically the relaxation spectrum for a Monte Carlo simulation in a 1D potential with many local minima. We take the potential: 
\begin{align}
U(x) \, = \, x^2/2 + 3 \, \sin 9 x
\label{eqmanymin}
\end{align}
inside a box $x \in (-5,5)$. This potential has many local minima as illustrated in the left panel of Fig.~\ref{FigNmin}. The eigenspectrum (see Fig.~\ref{FigNmin} right panel)  features a localization transition at $a^* \simeq 2.1$  as in the prototype cases with only a single minium. At variance with the spectrum for $U(x) = x^2/2$ (see Fig.~1 from the main text), many quasi-degenerate eigenvalues are present near $\lambda=1$, for low values of the jump amplitude $a$. In this regime indeed, hopping over the barrier is thermally activated and the mimima become metastable. 

%\newpage 

\end{appendix}

%\showmatmethods{} % Display the Materials and Methods section

%\acknow{Please include your acknowledgments here, set in a single paragraph. Please do not include any acknowledgments in the Supporting Information, or anywhere else in the manuscript.}

%\showacknow{} % Display the acknowledgments section

% Bibliography
\bibliography{mainbib.bib}

\begin{thebibliography}{53}
\bibitem [{Buf()}]{Buffon}%
  \BibitemOpen
  \href@noop {} {}\bibinfo {note} {When a needle of length $\ell$ is randomly
  tossed over a floor with parallel equidistant strips of size $2\ell$, the
  probability that the needle crosses a line between two strips is
  $1/\pi$}\BibitemShut {NoStop}%
\end{thebibliography}

\end{document}